\documentstyle[pptwocol,aaspp4,astrobib]{preprint}

\newcommand{\ltappeq}{\raisebox{-0.6ex}{$\,\stackrel
{\raisebox{-.2ex}{$\textstyle <$}}{\sim}\,$}}
\newcommand{\gtappeq}{\raisebox{-0.6ex}{$\,\stackrel
{\raisebox{-.2ex}{$\textstyle >$}}{\sim}\,$}}

\begin{document}

\title{\bf{HST/FOS Eclipse Observations of the Nova-like Cataclysmic
Variable UX Ursae Majoris\footnote{Based on observations with the
NASA/ESA {\em Hubble Space Telescope}, obtained at the Space Telescope
Science Institute, which is operated by the Association of
Universities for Research in Astronomy, Inc., under NASA contract
NAS~5-2655}}}

\author{Christian Knigge, Knox S. Long}
\affil{Space Telescope Science Institute, 3700 San Martin Drive,
Baltimore, MD 21218 \\ knigge@stsci.edu, long@stsci.edu}

\author{Richard A. Wade} 
\affil{The Pennsylvania State University,
Department of Astronomy and Astrophysics, \\ 525 Davey Laboratory,
University Park, PA 16802 \\ wade@astro.psu.edu}

\author{Raymundo Baptista} 
\affil{Departamento de Fisica, Universidade
Federal de Santa Catarina, Campus Universitario, Trindade, 88040
Florianopolis, Brasil \\ bap@fsc.ufsc.br}

\author{Keith Horne} 
\affil{Department of Physics and Astronomy, The
University of St. Andrews, North Haugh, St. Andrews, Fife, KY16 9SS,
UK \\ kdh1@st-andrews.ac.uk}

\author{Ivan Hubeny} 
\affil{NASA Goddard Space Flight Center,
Greenbelt, MD 20771 \\ hubeny@stars.gsfc.nasa.gov}

\author{Ren\'e G.M. Rutten} 
\affil{Isaac Newton Group, Apartado de
correos 321, E-38780 Santa Cruz de La Palma, Spain \\ rgmr@ing.iac.es}

\begin{abstract}

We present and analyze {\em Hubble Space Telescope} observations
of the eclipsing nova-like cataclysmic variable UX~UMa obtained
with the {\em Faint Object Spectrograph}. Two eclipses each were
observed with the G160L grating (covering the ultraviolet 
waveband) in August of 1994 and with the PRISM (covering the
near-ultraviolet to near-infrared) in November of the same year. The
system was~$\sim$50\% brighter in November than in August, which, if
due to a change in the accretion rate, indicates a fairly substantial
increase in $\dot{M}_{acc}$ by $\gtappeq 50\%$.

The eclipse light curves are qualitatively consistent with the gradual
occultation of an accretion disk with a radially decreasing
temperature distribution. The light curves also exhibit asymmetries
about mid-eclipse that are likely due to a bright spot at the 
disk edge. Bright spot spectra have been constructed by differencing 
the mean spectra observed at pre- and post-eclipse orbital 
phases. These difference spectra contain ultraviolet absorption lines
and show the Balmer jump in emission. This suggests that part of the
bright spot may be optically thin in the continuum and vertically 
extended enough to veil the inner disk and/or the outflow from UX~UMa
in some spectral lines.

Model disk spectra constructed as ensembles of stellar atmospheres 
provide poor descriptions of the observed post-eclipse spectra, 
despite the fact that UX~UMa's light should be dominated by 
the disk at this time. Suitably scaled single temperature model stellar
atmospheres with $T_{eff} \simeq 12,500-14,500$~K actually provide a
better match to both the ultraviolet and optical post-eclipse
spectra. Evidently, great care must be taken in attempts to derive
accretion rates from comparisons of disk models to observations.

One way to reconcile disk models with the observed post-eclipse
spectra is to postulate the presence of a significant amount of
optically thin material in the system. Such an optically thin
component might be associated with the transition region
(``chromosphere'') between the disk photosphere and the fast wind 
from the system, whose presence has been suggested by
Knigge~\&~Drew~(1997). In any event, the wind/chromosphere is likely
to be the region in which many, if not most, of the UV lines are
formed. This is clear from the plethora of emission lines that appear
in the mid-eclipse spectra, some of which appear as absorption
features in spectra taken at out-of-eclipse orbital phases.

\end{abstract}

\keywords{accretion, accretion disks --- binaries: close --- novae,
cataclysmic variables --- stars: individual (UX~UMa) --- ultraviolet:
stars}

\section{Introduction}
\label{introduction}

Nova-like (NL) cataclysmic variables (CVs) provide a unique
opportunity to study the process of disk accretion in what is likely
to be its simplest form. These systems are semi-detached binary stars,
in which mass is transferred from a Roche lobe filling, low-mass,
late-type secondary onto an accretion disk around a mass-gaining white
dwarf (WD) primary. NLs have relatively constant light curves, unlike
another class of disk-accreting CVs, the dwarf-novae (DNe), which
undergo 3-5 mag outbursts on time scales of weeks to months. The
difference between NLs and DNe is probably related to the rate at
which mass is transferred from the secondary onto the disk
\cite{meyer2,smak3}. In DNe, this accretion rate is below some
critical value, and the disk oscillates between a low-viscosity,
optically thin, quiescent state and a high-viscosity, optically thick,
outburst state. In NL variables, on the other hand, the accretion rate
is higher, and the disk is always in the optically thick high
state. Since accretion disk theory should be on firmest ground when
steady-state, optically thick conditions may reasonably be assumed,
observations of NLs provide an excellent testbed for our basic
understanding of accretion disk physics.

Observations of eclipsing systems are particularly important in this 
respect because phase-resolved light curves around orbital
phase $\phi_{orb}=0.0$ (when the secondary star is in
front of the disk and WD as seen from Earth) contain information about
the spatial brightness distribution of the accretion disk and other
system components (e.g. \citeNP{horne3}). Here, we present eclipse
observations of the NL variable UX~UMa obtained with
the {\em Faint Object Spectrograph} (FOS) on the {\em Hubble Space
Telescope} (HST). The data cover four primary eclipses with high time
resolution and extend over a total wavelength range of nearly
7000~\AA. The purpose of the present paper is to describe the
observations and to determine the spectral properties of the system 
components that can be isolated in the time-resolved spectra (the
accretion disk, the uneclipsed light and the bright spot). We have
also detected coherent 29-s dwarf nova type oscillations in some of
the data, which will be analyzed in a companion paper (\citeNP{me7},
hereafter Paper~II; see Warner \& Nather 1972 and Nather \& Robinson
1974 for previous detections of UX~UMa's 29-s
oscillations\nocite{warner4,nather1}). Eclipse mapping studies based
on this data set, using both direct model fitting and image
reconstruction techniques, are also being carried out and will be
presented separately (Knigge~et al. 1998, hereafter Paper~III;
Baptista~{et al.}~1998, hereafter Paper~IV; both in preparation)

\section{Observations}
\label{observations}

The target of our observations, UX~UMa, is a bright ($V \simeq 12.8$),
eclipsing nova-like variable with an orbital period of 4.72 hrs
\cite{kukarkin1}. Its system parameters have recently been derived by
\citeN{baptista1}, and we will not attempt to refine them here. 
Instead, we list Baptista~{et al.}'s values in
Table~\ref{tbl-systempars} for reference and will assume them
throughout the paper unless explicitly noted otherwise.

UX~UMa was observed with the FOS onboard the HST in August and
November of 1994. For the August observations, the G160L grating on
the BLUE digicon (nominal wavelength coverage 1140~\AA~-~2508~\AA) was
used with the {\bf 1.0} (0.86''~diameter) aperture; in November,
the PRISM on the RED digicon (nominal wavelength coverage
1850~\AA~-~8950~\AA) was used 
with the {\bf 4.3} (3.66''~$\times$~3.71'') aperture. This
combination of gratings gives the widest overall wavelength coverage
that was obtainable with the FOS, at the price of relatively low spectral
resolution. The spectral resolution of the G160L
was 6.6~\AA~FWHM with our instrumental set-up, and that of the PRISM,
which has a highly non-uniform dispersion, varied from about
4~\AA~near the short wavelength end to more than 400~\AA~at the
longest wavelengths. Two eclipses were observed in both epochs. The
two August sequences covered consecutive orbital cycles, but the 
November sequences were separated by two unobserved eclipses. All
observations were carried out in RAPID mode, with a new exposure
beginning approximately every 5.4~s. A log of the four observing
sequences is given in Table~\ref{tbl-log}.

All of the observations occurred normally, with the target well
centered in the aperture. The data were reduced using the standard
STSDAS pipeline software with the set of calibration files that was
available in 1996 May. STSDAS pipeline reduction includes
flat-fielding corrections, background subtraction, as well as
wavelength and absolute flux calibrations. To avoid having to apply
additional and uncertain corrections for geocoronal, dark current and
second order contributions, we have restricted our analysis of the
data to somewhat smaller than nominal wavelength regions, namely
1230~\AA~--~2300~\AA~for the G160L spectra and
2000~\AA~--~8000~\AA~for the PRISM spectra. The photometric accuracy 
of the reduced spectra should be about 4\%.

In observations with the G160L grating, the zeroth order
undispersed light is also recorded and can be used to construct a
broad-band optical/UV light curve. The zeroth order light has a
bandpass with full-width at half-response of 1900~\AA~and a pivot
wavelength of 3400~\AA. The G160L zeroth order photometry has so far
only been flux-calibrated against pre-COSTAR observations
\cite{horne5,erac2}, but \citeN{erac2} do predict a post-COSTAR
response of 950~cts~s$^{-1}$~mJy$^{-1}$ ($\simeq 3.7 \times
10^{17}$~cts~s$^{-1}$~[erg~cm$^{-2}$~s$^{-1}$~\AA$^{-1}$]$^{-1}$) for 
our instrumental set-up. This calibration is, however, not very
precise, as even the pre-COSTAR response is only accurate to about
50\% \cite{erac2}.

Continuum light curves were constructed from the first order data for 
a number of wavelength bands as shown in
Figures~\ref{fig-g160lcontlights}~and~\ref{fig-prismcontlights}.
These bands cover the observed spectral range but avoid obviously
line-blanketed regions and strong emission/absorption features. The
phase-averaged mid-eclipse spectra (see below) proved to be
particularly useful in selecting suitable continuum windows. Each
light curve was computed by averaging the flux within a given
continuum band for each spectrum in a given observing sequence and
phasing the data according to the linear ephemeris given by
\citeN{baptista1}. In addition to these continuum light curves,
Figures~\ref{fig-g160lcontlights}~and~\ref{fig-prismcontlights} also
show high S/N, ``white-light'' light curves that have been constructed
as time series of the average fluxes over the full adopted wavelength
range of each grating. For the two G160L observing sequences, the
zeroth order light curves, calibrated according to the response
estimate given above, are also plotted in
Figure~\ref{fig-g160lcontlights}.

The spectral properties of the various system components in UX~UMa are
most easily investigated by relying on mean spectra obtained by
averaging over suitable orbital phase ranges. For the present study,
mean pre-, mid- and post-eclipse spectra are required, which have been
calculated as averages over the following phase ranges
\begin{trivlist}
\item[(1)] Pre-eclipse: $\phi<0.95$
\item[(2)] Mid-eclipse: $0.98<\phi<1.02$
\item[(3)] Post-eclipse: $\phi>1.04$
\end{trivlist}
These spectra are shown in
Figures~\ref{fig-g160lavespecs}~and~\ref{fig-prismavespecs}, which
also give suggested identifications for the strongest emission and
absorption lines in the spectra.

Since the shape and apparent magnitude of UX~UMa's spectrum could be
affected by interstellar reddening, we have inspected all mean spectra
for evidence of the characteristic 2175~\AA~dust absorption
feature. None was found, consistent with the non-detection of the same
reddening indicator in an analysis of IUE spectra
\cite{verbunt3}. Consequently, we adopt E(B-V)=0.0 as our preferred
estimate, but will consider values as high as Verbunt's upper limit of
E(B-V)=0.04 where our conclusions might otherwise depend on this.

\section{Analysis of Light Curves}

\subsection{General Characteristics}
\label{general}

Overall, the light curves in
Figures~\ref{fig-g160lcontlights}~and~\ref{fig-prismcontlights} are
typical of eclipsing NL variables and DNe in outburst
(e.g. \citeNP{rutten1}). Specifically, from short to long wavelengths,
the eclipses become increasingly broad and shallow, and the
out-of-eclipse flux level drops off. This is consistent with the
eclipse of an accretion disk with a relatively blue spectrum (see
Section~\ref{prepost}) and a temperature distribution that decreases
with radius. Substantial short time-scale variations unrelated to the
eclipse -- ``flickering'' -- are also seen. This, too, is common among
NLs and DNe.

All of the light curves are asymmetric. At wavelengths below the 
Balmer jump at 3646~\AA, the asymmetry manifests itself primarily as 
a difference between the pre- and post-eclipse flux levels. Thus, even
though the deep eclipse of the central parts of the disk appears to be
over at orbital phases $\phi \gtappeq 1.06$, the fluxes have
recovered to just 50\%~--~80\% of the pre-eclipse level at this
point. Beyond this, the post-eclipse flux at these wavelengths remains
roughly constant to the end of the observing
sequences. At longer wavelengths, the asymmetry is perhaps better
described as an extended eclipse egress relative to ingress. However,
even at these wavelengths the  pre-eclipse flux level is never fully
recovered.

Asymmetries of this type could result from orbital phase-dependent
absorption in the system. More conventionally, they tend to be
attributed to a ``bright spot'' (BS) produced by the impact of the
accretion stream 
from the secondary onto the disk edge. Since the stream leads the
secondary, the BS 
should be located in the quadrant of the disk between phases 0.75 and
1.0. If (part of) the BS can be described as a spot painted on the
outside of the disk rim, it will be visible for about half an orbital
cycle. During this time, its uneclipsed light curve should be roughly
sinusoidal, with the maximum occurring when the spot is facing the
observer. The BS can also be eclipsed by the secondary, but due to its
position at the disk edge and off the line of centers of the two
stars, mid-eclipse for the BS will be delayed with respect to the time
of conjunction.
        
Most of the characteristic features of the light curves in
Figures~\ref{fig-g160lcontlights}~and~\ref{fig-prismcontlights} can be
interpreted within this framework. The failure of all light curves to
recover their pre-eclipse flux levels after conjunction implies that
the BS eclipse lasts until close to the end of the observing runs
and/or that the BS light is near maximum at pre-eclipse phases, so
that its contribution at post-eclipse phases is diminished. The difference
between the light curve morphologies at short and long wavelengths can
also be understood qualitatively. If the BS is relatively compact,
the width of its eclipses will be the same at all wavelengths. Disk
eclipses, on the other hand, will widen at 
longer wavelengths, where the cool outer parts of the disk contribute 
relatively more to the flux. A superposition of spot and disk light
curves might therefore be able to reproduce the observed light curve
shapes. We will test this interpretation quantitatively in Paper III.

Apart from short time scale variations away from eclipse, the light
curves constructed from Runs~1 and 2 (G160L) are similar, as are
the Run~3 and Run~4 (PRISM) light curves. The main exception is the
longest wavelength (7000~\AA~--~8000~\AA) PRISM light curves, which
show an $\sim 20$\% flux decrease between Runs 3 and 4.  Since only
the red end of the spectrum is affected this brightness change may be
related to a change in the cool, outer parts of the disk.
%To give a simple numerical example, a reduction of the outer disk radius 
%by approximately 25\% reduces the flux in these wavelength regions by 
%the required amount 
%in the optically thick, steady-state disk model
%spectra described in Section~\ref{spectrum_modeling} 
%(assuming a plausible 
%accretion rate of $6 \times 10^{17}$~g~s$^{-1}$). However, since
%the eclipse widths of the long wavelength PRISM light curves for Runs
%3 and 4 are not very different, this particular interpretation should 
%not be taken too literally. 

The continuum light curves between 2000~\AA~and~2300~\AA, which are
common to the G160L and PRISM observing sequences, show that UX~UMa
was about 50\% brighter in November than in August.  If interpreted as
reflecting a change in the accretion rate, a 50\% flux increase in
this wavelength region corresponds to an increase in $\dot{M}_{acc}$
by $\gtappeq 50$\% in the optically thick disk model spectra described
in Section~\ref{spectrum_modeling}, for all accretion rates $\gtappeq
10^{17}$~g~s$^{-1}$. From the overlap region of the pre- and
post-eclipse spectra described in Section~\ref{spectrum_modeling}, we 
find that the ratios of the pre- and post-eclipse levels in November
($\simeq 1.2$--$1.4$) were similar, although slightly larger than
those in the August data ($\simeq 1.4$--$1.7$). Thus the BS and the
disk appear to have brightened roughly in step, with the BS being
somewhat more prominent in November. This behaviour is qualitatively
consistent with an increase in the rate of mass supply from the
secondary. 

The UV flux level in our August/G160L data is below that seen in
earlier HST/GHRS observations of UX~UMa, and the system was somewhat
fainter than usual even during those observations \citeN{mason1}. Thus 
the brighter state the system was in during the November observations
with the PRISM is probably more representative of the average
mass transfer rate in UX~UMa. Brightness changes of the magnitude we
observe are not uncommon among NLs, many of which exhibit occasional
high and low states (e.g. \citeNP{dous1}).

Based on the long wavelength light curves of Runs~3 and~4, we estimate
that the half-width of the full disk eclipse is $\Delta \phi \simeq
0.1$. Using the geometric method of Sulkanen, Brasure \&
Patterson (1981)\nocite{sulkanen1}, this yields an estimate for the
disk radius of $R_{disk} \simeq 35 R_{WD}$ when combined with
Baptista~{et al.}'s (1995) system parameters for UX~UMa. No useful
disk radius estimate can be made from the G160L light curves, since
the outer disk contributes so little to the total light at UV
wavelength. We will analyze the eclipse geometry, including
constraints on the disk radius, in more detail in Paper~III. For the
modeling of disk spectra carried out in the present study
(Section~\ref{spectrum_modeling}), the estimate just quoted is
sufficiently accurate.

\subsection{Eclipse Timings}
\label{ominusc}

Figures~\ref{fig-g160lcontlights}~and~\ref{fig-prismcontlights} show
that Baptista~{et al.}'s (1995) ephemeris predicts the times of
minima in 1994 August and November quite well. To quantify this
agreement, we have measured accurate eclipse timings from the first
order white-light light curves in two ways. First, a
parabolic function was fitted to the central part of the eclipse. This
gives an estimate of the {\em time of minimum} in a given light
curve. Second, a numerical derivative was constructed from a slightly
smoothed version of each light curve and inspected for extrema. The
corresponding orbital phases should correspond to the ingress/egress
phases of the WD and/or disk center, and their mid-point gives an
estimate of {\em mid-eclipse times} (times of conjunction). We
estimate that the internal error on both types of O-C measurements is 
less than 0.001 cycles. However, their reliability as estimates of
times of conjunction could be subject to somewhat larger systematic
uncertainties because the BS, whose eclipse will not be centered on
$\phi=0$, also contributes to the total light from the system.

Our new eclipse timings for UX~UMa are listed in
Table~\ref{tbl-timings}. We find that (a) there are no obvious
differences between the O-C values derived for the August (Runs 1 and
2) and November (Runs 3 and 4) data sets with a given method, and (b)
the O-C values deduced from the parabolic fit to eclipse center, which
have a mean of +0.0033 cycles, are systematically larger than those
derived from the derivative light curve, which have a mean of +0.0024
cycles. The difference is probably due to the effect of the
BS on the light curves. Since the BS eclipse is centered on some phase
after conjunction, it will tend to skew measurements of eclipse
timings towards later phases. This suggests that the O-C values
derived from the derivative light curve are more reliable estimates of
times of conjunction, as expected.

We have not tried to refine Baptista~{et al.}'s (1995) ephemeris on
the basis of our new eclipse timings, since the epochs of the new
measurements add little to the long timeline of observations on which
Baptista~{et al.}'s (1995) ephemeris is based. The measured O-C values
are in any case quite small by UX~UMa's standard (c.f. Baptista~{et
al.}'s Figure~5).

\section{Analysis of Spectra}
\label{spectrum_modeling}

\subsection{Method}

In a first attempt to characterize and interpret the spectra of the
system components that can be isolated in the data, we have fit two
types of (optically thick) models to the observations:

\begin{trivlist}
\item[(1)] a single (effective) temperature, solar abundance stellar
atmosphere (SA) model (parameters: temperature, gravity, normalization
[projected area/distance$^2$])
\item[(2)] a standard, steady-state accretion disk radiating as an
ensemble of solar abundance SAs (parameters: accretion rate, disk
radius, normalization [distance])
\end{trivlist}
The single temperature SA model spectra were calculated by
interpolation in temperature and gravity from a suitable grid of
(angle-averaged) SA spectra that were generated from Kurucz (1991)
{\sc atlas} \nocite{kurucz2} structure models using Hubeny's spectral
synthesis code {\sc synspec} \cite{hubeny2}.

The disk models were constructed as area weighted sums of SAs over the
face of the accretion disk. In this summation, the standard relation
for the run of effective temperature with radius in a steady-state
accretion disk was assumed \cite{pringle1}, i.e.:
\begin{equation} 
        T_{eff}(R) \; = \; T_{*} \; { \left( \frac {R_{WD}} {R}
                \right)}^ {3/4} \; {\left( 1 - \sqrt{ \frac
                {R_{WD}}{R} } \right)}^{1/4},
\label{eq-disk}
\end{equation} 
where $R$ is the distance from the WD on the disk and $R_{WD}$ is the
WD radius. $T_{*}$ is a function of the mass accretion rate,
$\dot{M}_{acc}$, and the WD mass and radius:
\begin{equation} 
T_* = \left(\frac{3GM_{WD} \dot{M}_{acc}}{8 \sigma \pi
R_{WD}^{3}}\right)^{1/4},
\label{eq:tstar}
\end{equation} 
with $\sigma$ and $G$ being the Stefan-Boltzmann and Gravitational
constants, respectively. For the local disk gravity, the approximate
relation given by \citeN{herter1} was adopted. In all disk models, we
fixed the disk radius at $R_{disk} \simeq 35 R_{WD}$, as derived in
Section~\ref{general} from the PRISM light curves. Given the
three month time delay between the first two and last two
observing sequences, this estimate may not be appropriate for the
G160L data. However, since the cool, outer disk contributes little to
the total flux at UV wavelengths, disk model fits to the G160L spectra
are insensitive to the adopted disk radius.

Considering the large wavelength range that is covered by the
observations, we decided to account at least approximately for
wavelength-dependent limb darkening in our disk
models. To this end, we used {\sc synspec} to calculate the specific
intensities for an angle equal to UX~UMa's inclination of i=71$^o$
from a suitable subset of the Kurucz {\sc atlas} structure models and
used these spectra, rather than ones corresponding to angle-averaged
fluxes, in the construction of the disk model spectra. As shown by
\citeN{diaz1} for the UV wavelength region -- where limb-darkening is
strongest -- this should be a reasonable approximation for the
optically thick disk models that are appropriate to UX~UMa. To allow
consideration of very high mass accretion rate models -- in which the
maximum effective temperature can exceed 50,000~K (the highest
temperature for which Kurucz structure models exist) -- we also
supplemented this grid of specific intensities with several spectra
that were generated by {\sc synspec} from {\sc tlusty}
\cite{hubeny1,hubeny3} structure models with $50,000~K \leq T_{eff}
\leq 140,000~K$.

Since the size of the emitting source is fixed in the disk models, 
the normalization constant required to match the observed flux 
measures the implied distance of the system. Distance estimates 
for UX~UMa in the literature vary between 216~pc and 345~pc
(see \citeNP{rutten1} and \citeNP{baptista1} for overviews) with the
most recent determination falling at the high end of this range
($345\pm34$~pc; Baptista {\em et al.}~1995).
%\footnote{The low estimate of
%216~pc was derived by \citeN{bailey1} on the basis of Frank~{\em et
%al.}'s (1981) data and model for the system by using the K-band
%surface brightness of the secondary star as a distance
%indicator. Frank~{et al.}'s (1981) own estimate of the distance
%was $340\pm110$~pc. An attempt to reproduce Bailey's value, using the
%same data, method and model, resulted in a somewhat larger distance
%prediction of about 250~pc. Essentially the same number was obtained
%when Ramseyer's (1994) \nocite{ramseyer1} recent recalibration of
%Bailey's K-surface brightness relationship for late type dwarf stars 
%was taken into account.} 
Adopting the latter, formally most accurate estimate as the preferred
distance towards the system, but allowing the uncertainty in this to
be large enough to include all other previous distance determinations,
we conservatively consider as viable all disk models for which the
implied distance lies within the range 200~-~500~pc.

Neither of our two types of models is expected to reproduce the strong
absorption and emission lines in UX~UMa's UV spectrum, as these are
most likely formed in the accretion disk wind from the system or in a
transition region (``chromosphere'') between the disk and the outflow
(\citeNP{drew5,prinja3,me6}; see also Section~\ref{mid}).
Consequently, wavelength regions containing strong lines were excluded
in model fits to the G160L observations. All model spectra were
smoothed to the appropriate instrumental resolution before comparing
them to the data. In the following sections, we describe the results
of these comparisons for Runs~1 (G160L) and 3 (PRISM). Results for
Runs~2 (G160L) and 4 (PRISM) were essentially identical, as expected,
since the observed spectra themselves are so similar (see
Figures~\ref{fig-g160lavespecs}~and~\ref{fig-prismavespecs}).

\subsection{The Spectrum of the Accretion Disk} 
\label{prepost}

The average pre- and post-eclipse spectra should be dominated by disk
light. The post-eclipse spectra, in particular, should represent the
isolated spectrum of the accretion disk well, since the BS contributes
to the pre-eclipse flux, but is likely to be occulted at most or all
post-eclipse orbital phases (see Section~\ref{general}). We therefore
use the post-eclipse spectra as our best estimates of the 
spectrum of UX~UMa's accretion disk in this section and fit them with
both disk models and single temperature SAs. The SA model fits are
useful as a convenient parameterization of the observed spectral shape
and as benchmarks against which the goodness-of-fit achieved by the
physically more relevant disk models may be judged.
%\footnote{We note in passing that, since the disk radius
%is kept fixed in the disk models, they actually contain one 
%less degree of freedom than do the single temperature SAs. However, 
%this formal difference in the complexity of the two prescriptions is
%at best marginally significant, since one of the SA model parameters
%(gravity) usually exerts much less leverage on the fits than the other
%two (temperature and normalization).}

The best fitting disk and SA models are shown along with the Run~1
and~3 post-eclipse spectra in Figure~\ref{fig-postfits}. While the
accretion rates implied by the disk models -- $1-2 \times 10^{17}$
g~s$^{-1}$ -- are not unreasonable {\em a priori}, the match to the
data achieved by these models is not particularly good. Moreover, the
distance implied by both ``best'' disk model fits is 200~pc, i.e. 
equal to the lower limit that we imposed in the
least-squares minimization. Similar results were obtained in disk
model fits to the mean pre-eclipse spectra and to the 
corresponding spectra in Runs~2 and 4. However, the most
disconcerting aspect of Figure~\ref{fig-postfits} is that
the disk models fit the data worse than single temperature SAs.  In
fact, the best-fitting SA models -- which have $T_{eff} \simeq
12,500-14,500$~K -- provide a remarkably good match to the data at all
observed wavelengths away from the Balmer jump, including the UV. (See
Dickinson~{et al.}~[1997]\nocite{dickinson1} for a similar finding for
the NL CV V795~Her.) We have also tried
to fit disk model spectra constructed as area-weighted sums of
blackbodies to the pre- and post-eclipse spectra, but these, too, fail
to improve upon the fits achieved by single temperature SA models.

According to Rutten~et~al.~(1994)\nocite{rutten2}, the spectral type
of the secondary in UX~UMa lies in the range by K7 -- M0. To
illustrate how the observed spectra (and hence the model fits) may 
be affected by the presence of this additional system component, we
have included in Figure~\ref{fig-postfits} SA model spectra for
K7 and M0 main-sequence stars from the \citeN{buser1} compilation of
{\sc atlas} SA models (our grid of {\sc synspec} generated models does
not extend to the required low effective temperatures.)
To allow a comparison with the disk model fit, the representative 
secondary star spectra in Figure~\ref{fig-postfits} have been scaled 
to the flux expected from the secondary if the distance to the system 
were 200~pc. The scaled secondary star models
in Figure~\ref{fig-postfits} show that the contamination of the
post-eclipse spectra by light from the secondary is likely to be
small, except perhaps at the longest optical wavelengths. There, the
disk model spectrum already falls off more slowly with wavelength than
the Run~3 post-eclipse spectrum, so that subtraction of the very red
secondary star from the data prior to fitting the disk models is not
likely to improve the fit.

Concerned that the apparent failure of the disk models may be caused 
by reddening or by some other flaw in our model fitting procedure
(perhaps related to the selection of continuum 
windows or to the highly non-uniform dispersion of the PRISM), we
decided to compare these models against the data in another way. 
Thus, in each panel of Figures~\ref{fig-dred1fits}~and~\ref{fig-dred3fits},
three disk model spectra are plotted alongside the observations, 
with each model shown normalized to the data at a preselected
continuum wavelength (1450~\AA~for the G160L data, 4500~\AA~for the PRISM
data). The accretion rates corresponding to the models in each panel
have been selected so as to make the implied distances for this choice
of normalization completely cover the allowed range. Also, results are
shown in Figures~\ref{fig-dred1fits}~and~\ref{fig-dred3fits} for both
minimal (E(B-V)=0.00, top panels) and maximal (E(B-V)=0.04, bottom
panels) assumptions about the reddening towards UX~UMa.

It is clear from Figures~\ref{fig-dred1fits}~and~\ref{fig-dred3fits}
that the mismatch between the data and the best-fitting disk models is
real and cannot be attributed to the effects of reddening or explained
as an artifact of the adopted fitting procedure. For example, even
though a disk model with $\dot{M}_{acc}=9\times10^{17}$~g~s$^{-1}$
does produce an improved fit to the spectral shape of the maximally
dereddened PRISM data, the Balmer jump in this model is still much too
large. Furthermore, the match of all disk models to the dereddened
G160L UV data is actually worse than that to the uncorrected
spectrum. This is because the G160L data 
cover the broad 2175~\AA~dust absorption feature. Thus, while
``dereddening'' does produce spectra that are bluer than the original
in most wavelength ranges, the effect is inverted at the longest UV
wavelengths ($\gtappeq$~1800~\AA) which are affected by the
2175~\AA~feature. As a result, the discrepancy between the disk models
and the data in that wavelength region is amplified, not reduced. It
is also worth noting from Figure~\ref{fig-dred1fits} that the UV
slopes of disk model spectra are actually highly insensitive to the
adopted accretion rate.

We may finally ask if the poor fits to the data achieved by the disk
models could be due to the contribution of the WD to the observed
spectra. While it is possible that the WD does contribute
non-negligibly to UX~UMa's UV flux \cite{baptista1}, it is unlikely
that adding a corresponding component to the disk model spectra could
improve the fit to the data in that wavelength region. This is because
only a hot WD can contribute significantly at all, which means that
any non-negligible WD spectral component must be very blue. This is
unlikely to be of much help in our modeling, since the disk models 
alone are already bluer than the observed UV spectra. This
intuitively plausible prediction has been borne out quantitatively in
previous analyses of UV CV spectra \cite{long2,me5}. Moreover, even a
hot WD would not contribute much to the flux at optical wavelength, so
the neglect of a WD component in our disk model fits to the PRISM data
is almost certainly unimportant. From a broader perspective, it is of 
course nevertheless very important to try to isolate the WD component
in our UV data. However, this is best achieved by means of careful,
quantitative analyses of the eclipse light curves
(c.f. \citeNP{baptista1}), rather than by examining the phase-averaged
spectra. The results of two such light curve analyses will be presented 
in Papers~III and IV.

Based on Figures~\ref{fig-postfits}~-~\ref{fig-dred3fits} and the
discussion up to now, we conclude that steady-state SA disk models 
are unable to describe the data acceptably and in fact fare worse in
this respect than simple, single temperature model SAs. This
conclusion appears to hold for all disk models with 
accretion rates that would retain consistency with existing distance
estimates towards the system. This need not imply, of course, that the
emitting region(s) in UX~UMa are better described {\em physically} by
a single temperature SA than by a steady-state accretion disk
model. In fact, as noted in Section~\ref{general}, the wavelength
dependence of the eclipse shape and depth does suggest a radially
decreasing effective temperature distribution, as expected in the
standard accretion disk picture. However, given that simple models
based on steady-state disk theory do not provide a good match to the
spectral shape of real accretion disk systems -- not even relative to
single temperature SAs -- great care must be taken in attempts to
infer disk parameters on the basis of any such models.

As an aside, we note that this caution is as relevant to eclipse
mapping studies, including those based on image reconstruction
techniques, as it is to studies relying directly on model fits to
observed accretion disk spectra. The reason for this is that the
accretion rate, $\dot{M}_{acc}$, cannot be inferred directly from a
disk's surface brightness distribution; what is needed instead is the
radial run of (effective) temperatures across the disk. It is this
intermediate step -- the transformation of surface brightness into
temperature -- which requires the specification of an emissivity law
(as well as a distance estimate). In practice, BB emissivities are
usually assumed at this stage. Thus, even though disk images obtained
with the maximum entropy method, for example, are independent of any
assumptions about the spectral properties of the disk, the radial
temperature distributions and accretion rates derived from such images
are not. It should also be kept in mind that both the shape and
normalization of radial temperature distributions derived in this way
are sensitive to errors in the adopted distance estimates.

Returning to the disk model fits, we note that the two main failings
of the models are (1) their excessively blue color at UV wavelengths
and, (2) the magnitude of the predicted Balmer jump, which is much
larger than observed.

Both of these problems have long histories. \citeN{wade1} first noted
that SA disk model spectra were systematically bluer than the spectra
of NL CVs observed with the {\em International Ultraviolet Explorer}
(IUE). He found that the problem could only be alleviated if
uncomfortably low accretion rates were adopted, which in turn produced
a large UV flux deficit in the model spectra for the accepted
distances towards the systems in his sample. The same problem has
since been encountered in analyses of (HUT) observations of 
high-state, non-magnetic CVs
\cite{long1,long2,me5}. Thanks to HUT's coverage of wavelengths all
the way down to the Lyman limit, the latter studies were able to show
more conclusively that models with low accretion rates do not offer a
viable solution, since they fail to fit the observed spectra shortward
of their turnover at about 1000~\AA. Similarly, the discrepancy
between observed and predicted Balmer jumps is also not
new. \citeN{wade4} and \citeN{dous2}, for example, have also 
constructed SA disk model spectra covering the optical and UV
ranges. Their models, as ours, were found to produce much larger
Balmer jumps than are 
observed. We will discuss possible resolutions to these problems in
Section~\ref{adc}. 
%In the meantime, we stress again that accretion
%rate estimates provided by disk model fits to the shape of CV 
%spectra must be regarded as unreliable.

\subsection{The Spectrum of the Uneclipsed Light}
\label{mid}

UX~UMa's eclipses are not total: some residual flux emerges at all UV
and optical wavelengths even at mid-eclipse (see
Figures~\ref{fig-g160lcontlights}~and~\ref{fig-prismcontlights}).
Some of this uneclipsed light is due to the secondary star, but the
outer parts of the disk on the far side from the secondary also
contributes, since UX~UMa's accretion disk is never fully occulted. In
addition to these two components, there may also be other sources of
radiation which remain uneclipsed by virtue of their vertical
extent. For example, (some of) the spectral lines in the UV region are
almost certainly formed in UX~UMa's accretion disk wind (Knigge \&
Drew~1997), and part of the BS (see Section~\ref{preminuspost}) may
also have the required geometrical and emissive properties to
contribute to the uneclipsed light (the required vertical extent would
be $z_{BS} \gtappeq 18~R_{WD}$).

The mean mid-eclipse spectra derived from Runs~1 and 3 are shown along
with the best-fitting single temperature SA model spectra in
Figure~\ref{fig-midfits}. The G160L UV spectrum has a relatively flat
continuum and is dominated by emission lines. It is easy to show that
the secondary star contributes negligibly to the uneclipsed light in
this waveband for any plausible combination of distance and spectral
type. The UV continuum shape is nevertheless reasonably well described
by a single temperature SA with log~g~=~2.0 and $T_{eff} = 9200$~K. If
we take these parameters to be meaningful, they must be representative
of the physical conditions in the outer accretion disk. However, for
a distance towards the system of 345~pc, the normalization of
the SA model fit predicts that the emitting source has a projected
area, $A_{proj}$, whose minimum linear size is $l_{min} =
\sqrt{A_{proj}/\pi}=20~R_{WD}$. If we identify this source with the
uneclipsed part of the accretion disk, this size estimate must be
increased by a factor of $1/\sqrt{\cos i} = 1.75$ to $l_{min} =
35~R_{WD}$ due to foreshortening. This is equal to the disk radius
estimate derived from the PRISM data and therefore too large to
describe the fraction of the disk that remains unocculted at
mid-eclipse. Thus the identification of the uneclipsed UV light with
just the outer parts of an optically thick, geometrically thin disk 
is probably not viable, unless the distance towards UX~UMa is
significantly less than 345~pc. Of course, the outer disk may be
neither optically thick nor geometrically thin, and the accretion 
disk wind and/or parts of the BS region may also contribute to the
mid-eclipse UV continuum flux.

The classical CV wind lines, N~{\sc v}~1240~\AA, Si~{\sc iv}~1400~\AA~
and C~{\sc iv}~1550~\AA, are hardly occulted (see
Figure~\ref{fig-g160lavespecs}), as expected if they are formed in the
vertically extended outflow. It should be noted, however, that recent
HST/GHRS eclipse observations of the C~{\sc iv}~line in UX~UMa show
that the weak line eclipses are at least partly due to narrow {\em
absorption} features in the out-of-eclipse line profiles,
which {\em are} occulted during eclipses (Mason~{\em et
al.}~1995). Knigge~\&~Drew~(1997) modeled these data and showed
that the region contributing net absorption to the line profiles can
be plausibly identified with a relatively dense, slow-moving
transition region between the accretion disk and the fast wind (see
also \citeNP{baptista1} and Section~\ref{adc}).

\citeN{me5} used the same model to fit five strong UV resonance lines
in a HUT spectrum of the DN Z~Cam 
in outburst and showed that several other lines that are
typically seen in the UV spectra of high-state CVs might be produced
in the dense transition region predicted by this wind model.  This
idea is supported by our data, inasmuch as not just the classical wind
features, but just about all identifiable spectral lines -- some of
which appear in absorption away from eclipse (e.g. Si~{\sc
iii}~1265~\AA, C~{\sc ii}~1335~\AA, N~{\sc iv}~1718~\AA) -- become
clear emission features in the mid-eclipse G160L spectra.

Whether a spectral line that is formed in such a region will
appear as an absorption or an emission feature depends on how much of
the line-forming region is seen projected against the bright accretion
disk at the relevant inclination and orbital phase. In an eclipsing
system, such as UX~UMa, much of the disk is occulted at
mid-eclipse. Consequently, the only parts of the line-forming region
that are visible at that phase are those that do not lie ``in front
of'' the disk as seen from Earth. Inevitably, therefore, an emission
line spectrum will be observed in that case at orbital phase
$\phi=0.0$. If thermal line emission from (as opposed to scattering
in) such a transition region is sufficiently strong, the spectral
features formed within it may also appear in emission even when the 
disk does lie within our line of sight.

The Run~3 PRISM mid-eclipse spectrum in Figure~\ref{fig-midfits} is
also much redder than its out-of-eclipse counterpart. However, its 
continuum shape is 
not well described by a single temperature SA (the best-fitting model
has log~g~=~2.5 and $T_{eff} = 11300$~K). The main problem is that for
the relatively low temperature needed to fit the overall shape of the
spectrum, SA models predict the presence of an extremely strong Balmer
jump. On the other hand, a Balmer absorption edge is not visible in
the data (the jump may even be slightly in emission).

The temperature of the best-fitting SA model is still much higher than
expected for the secondary star. To estimate the likely contribution
of the secondary to the optical spectrum of the uneclipsed light, two
representative spectra are again shown in
Figure~\ref{fig-midfits}. The stellar radius and spectral types
adopted for these secondary star model spectra are the same as for
those in Figure~\ref{fig-postfits}, but the predicted fluxes have been
rescaled to correspond to our preferred distance estimate of
345~pc. For these parameters, the secondary contributes measurably to
the uneclipsed light above 4000~\AA, but does not dominate even at
the longest wavelengths. The shape of the observed mid-eclipse
spectrum also makes it clear that the secondary cannot be the sole
(and is probably not even the main) contributor to the uneclipsed
optical light. Instead, the absence of a Balmer jump in the spectrum
suggests that optically thin material may contribute to the mid-eclipse
spectrum, once again implicating the outer disk, the BS
and/or the accretion disk wind.

%We finally note that for distances significantly below our adopted
%lower limit of 200~pc ($\ltappeq$~170~pc in fact), the SA secondary
%star models for both spectral types shown in Figure~\ref{fig-midfits}
%exceed the observed mid-eclipse flux at the longest optical 
%wavelengths. If the secondary were an even cooler dwarf star with an
%effective temperature of, say, 3500~K (spectral type M2V), conflict
%with the absolute flux level at mid-eclipse could in principle be
%avoided for distances as low as about 130~pc.

\subsection{The Spectrum of the Bright Spot} 
\label{preminuspost}

If the light curve asymmetries in
Figures~\ref{fig-g160lcontlights}~and~\ref{fig-prismcontlights} are
due to the BS, its spectrum can be isolated 
approximately by taking the
difference between the average pre- and post-eclipse spectra. 
These difference spectra are shown in Figure~\ref{fig-spotfits} for
Runs~1 and 3, along with the best-fitting single temperature SA
models. 

The continuum shape of the UV BS spectrum is fairly similar to that of
the accretion disk (c.f. Figure~\ref{fig-postfits}) and thus has a
comparable color temperature (about 16,000~K). The minimum linear size
(calculated once again as $l_{min} = \sqrt{A_{proj}/\pi}$) provided by
the SA model fit to the G160L data is about 5~$R_{WD}$ for an assumed
distance of 345~pc. This is probably a reasonable estimate for
the size of the BS. We nevertheless caution that both temperature and
size estimates may be unreliable for reasons which will become clear
below.

All of the spectral lines that can be identified in the UV BS spectrum
appear to be in absorption (Si~{\sc iii}~1300~\AA, Si~{\sc
iv}~1400~\AA, C~{\sc iv}~1550~\AA~and Al~{\sc iii}~1860~\AA). This is
interesting, since at least two of the same features (Si~{\sc
iv}~1400~\AA~and C~{\sc iv}~1550~\AA) are in emission in the pre- and
post-eclipse spectra. Given that most of the UV lines are probably 
formed in some part of UX~UMa's accretion disk wind, one possibility is 
that some of the radiation emitted by the BS is scattered out of our  
line-of-sight by the outflow. In this picture,
the line-of-sight to the BS must pass through the wind. Alternatively,
the spot may be optically thick and produce an absorption
line spectrum similar to a SA. Finally, the spot may be 
(partially) optically thin in the continuum, and act as a
veiling curtain for the radiation emitted by the disk 
and/or the wind. In this case, our line of sight to the disk and/or 
wind must pass through at least part of the BS at pre-eclipse phases,
which requires the spot to have a significant vertical extent. 

The PRISM BS spectrum is even more striking. It, too, exhibits line
absorption features at near-UV wavelengths, but more importantly, the
Balmer jump is in emission. This suggests that the BS is
at least partially optically thin and strengthens the case for the veiling
interpretation of the spot's UV absorption line spectrum. It does not
confirm it beyond doubt for the G160L data, at least, because of the 
the significant change in brightness (accretion rate) that took place 
between August and November and which might have altered the
structure of the BS. In any case, given the signature of optically
thin material in the PRISM BS spectrum, the estimates of the BS size 
and temperature provided by the best-fitting SA model to these data
sets -- $l_{min} \simeq 5~R_{WD}$ and $T_{BS} \sim 2 \times 10^4$~K --
must be regarded with caution. This warning may also
apply to the estimates derived for the G160L data, if the BS region
was in fact also partially optically thin in August. 

We close this section by acknowledging that we cannot rule out that
the light curve asymmetries in our data are due to absorption
at post-eclipse orbital phases, rather than to the contribution of the
BS to the light at pre-eclipse phases. This possibility finds some
support in the fact that there are instances among existing
eclipse observations of UX~UMa (as well as among those of other
non-magnetic CVs) in which the ratio of pre- and post-eclipse flux
levels is $<1$ (see \citeNP{mason2} and references therein). If the
light curves asymmetries in our data are in fact due to absorption,
the difference spectra on which we have relied in this section are not
physically meaningful. However, in this case the ratio of post-eclipse
to pre-eclipse fluxes could be used to constrain the physical condition
in the absorbing material. This idea will be explored in more detail
in Paper~IV.

\section{Discussion}
\label{adc}

As noted in Section~\ref{prepost}, SA disk models, which should offer
the closest representation of a real accretion disk spectrum among the
models we have tried, do poorly in matching the observed
spectra of UX~UMa and other high-state CVs. Most importantly, the UV
colors predicted by SA disk models are too blue and the magnitude of
the Balmer jump is too large.

There are at least three possible reasons for these discrepancies.
First, the poor fits to the data may simply be due to the inadequacy
of our disk models. Certainly, given the very different geometries
involved, the atmospheres of real accretion disks may differ
considerably from those of ordinary stars. Some headway has already
been made in the construction 
of more self-consistent disk models, but the results so far are still
inconclusive. For example, \citeN{shaviv1} simultaneously solved the
(vertical) disk structure and radiative transfer equations and
presented the resulting self-consistent accretion disk model
spectra. These authors appeared to conclude that their models are able
to resolve the discrepancies between theory and
observations. \citeN{long2}, on the other hand, showed that similarly
self-consistent disk model spectra constructed on the basis of disk
model atmospheres calculated with Hubeny's code {\sc tlusdisk}
\cite{hubeny1,hubeny4}, differed little from those predicted by disk
models constructed as sums of ordinary SAs in the UV
waveband. Correspondingly, \citeN{long2} concluded that current
state-of-the-art disk models could not resolve the UV color
problem. The difference between these conclusions may simply be due to
the different wavelength ranges that were used in the comparisons with
observations, as well as perhaps to different views about what
constitutes success in the comparison of models and data.  However,
the somewhat different assumptions concerning viscosity and energy
release in the vertical direction that are adopted in {\sc tlusdisk}
and in the disk models calculations of Shaviv \& Wehrse (1991) may
also play a role. We finally note in this context that some
potentially important aspects of the disk physics -- such as the
radiative transport in the radial direction and the irradiation of the
disk by the WD and perhaps the boundary layer (BL) -- remain to be
explored.

A second way to account for the failure of SA disk models is to
question the assumptions made in the standard steady-state disk
model itself. For example, \citeN{long2} showed that the UV colors of
SA disk models could be reconciled with the observations if the hot,
inner parts of the disk (out to a few $R_{WD}$) were
removed. Physically, both the truncation of the disk by the magnetic
field of the WD -- a situation analogous to that in intermediate
polars -- or the removal of accretion energy by a powerful disk wind
might produce the observational signature of a ``missing'' (or cooler
than expected) inner disk. The former possibility has some additional
appeal in that it might explain the short period oscillations that
have been observed in UX~UMa and other non-magnetic CVs (see
Paper~II). However, one problem with this idea is that any type of
effective disk truncation may help to account for the observed UV
colors, but would probably worsen the Balmer jump discrepancy
(c.f. the spectra of disk annuli as a function of radius in la
Dous~[1989] and Shaviv \& Wehrse~[1991]).

A third possibility is that there are additional sources
of radiation in the system which must be included in model fits to the
obseved spectrum. However, as noted in Section~\ref{prepost}, any
additional optically thick components  -- e.g. due to the WD or BL --
are unlikely to resolve the UV color problem or to contribute
significantly at optical wavelengths. Thus, the only remaining option
is to identify the presumed additional emitting source with an
optically thin plasma \cite{hassall1}.

The idea that optically thin emission may contribute to the observed
spectrum in UX~UMa finds some empirical support in the fact that the
Balmer jump is strongly in emission in the PRISM spectra of the BS and
is either absent or perhaps very weakly in emission in the PRISM
mid-eclipse spectra. Independent ``theoretical'' support comes from
recent efforts to model the shapes, strengths and eclipse behavior of
the C~{\sc iv}~1550~\AA~resonance lines in the UV spectra of two
high-inclination NL CVs \cite{shlos2,me6}. Both of these studies
assumed that line formation occurs in a rotating, biconical accretion
disk wind and derived the required geometric and kinematic outflow
parameters from fits to the observed line characteristics.
Knigge~\&~Drew~(1997), in particular, modeled high quality HST/GHRS
eclipse observations of UX~UMa itself. A fundamental property of their
preferred wind model is very slow outflow acceleration close to the
disk plane, giving rise to an almost static, relatively dense ($n_e
\simeq 4 \times 10^{12}$~cm$^{-3}$) and vertically extended ($H \sim
10~R_{WD}$) ``transition region'' between the disk surface and the
fast-moving parts of the wind. Such a region -- which we shall call an
accretion disk chromosphere (ADC) hereafter, without meaning to imply any
direct correspondence with stellar chromospheres -- could in principle
produce significant amounts of optically thin recombination radiation.

To gain some insight into how much optically thin emission is required
to make a significant difference to the observed spectra, we have
followed \citeN{hassall1} in determining the emission measure (EM)
required to fill in the Balmer jump absorption in the disk model
spectra with Balmer jump emission produced by H~{\sc i} recombination
radiation. However, unlike \citeN{hassall1}, we have calculated the
required EM for a variety of plasma temperatures and accretion rates,
and also determined for each combination of disk model and H~{\sc i}
plasma the implied distance if this combination is to match the flux
of the observed Run~3 post-eclipse spectrum at 4500~\AA.

The results of this exercise are shown in
Figure~\ref{fig-thinstuff1}. It is easy to see from this that for
accretion rates $\dot{M}_{acc} \gtappeq 1\times10^{17}$~g~s$^{-1}$ and
distances in the range $200$~pc~$<d<500$~pc, emission measures in the
range $10^{56}$~cm$^{-3}$ $\ltappeq$ EM $\ltappeq 10^{58}$~cm$^{-3}$
and temperatures $T \ltappeq 4\times 10^5$~K are required. How
reasonable is this? Assuming a uniform, fully ionized ADC with the
parameters derived by Knigge \& Drew (1997), we find EM~$\simeq
n_e^2\times \pi R_{disk}^2 \times H = 5.7\times
10^{56}$~cm$^{-3}$. Regarding the temperature of the ADC,
Knigge~\&~Drew's~(1997) found that $T\simeq30,000$~K produced the
required amount of thermal C~{\sc iv} emission in their ``best-bet''
wind model. (It should be noted, however, that the C~{\sc iv}
ionization fraction in their wind model was not calculated, but simply
set to a plausible average value; also, collisional effects were
treated only approximately.)

The idea that optically thin ADC emission is contributing to the
observed spectra, and may be (partly) responsible for the poor fits
achieved by SA disk models, can be tested more directly by comparing
combined disk+ADC model spectra to the observations. To this end we
have used {\sc cloudy} \cite{ferland1} to calculate the emission
expected from a solar abundance coronal (i.e. collisionally ionized)
plasma at a variety of temperatures. For definiteness, $n_H=5\times
10^{12}$cm$^{-3}$ was assumed in all of these
models. Figure~\ref{fig-coronal} shows the resulting optically thin
model spectra.

Collisional equilibrium is probably not a very good approximation to
the conditions in the ADC, since photoionization by disk and perhaps
WD or BL photons is probably the dominant ionization process. However,
a photoionization model of the ADC would have to account in some way
for the 2-D character of the ADC geometry and its illumination. This
introduces complications that are unwarranted for our present, purely
exploratory calculations which are only intended to illustrate
qualitatively the effect of an optically thin ADC on the shape of the
emergent continuum. We note for reference that in the coronal models
Hydrogen (Helium) is essentially fully ionized for $T \gtappeq 2\times
10^4$~K ($T\gtappeq 8\times 10^4$~K). In a photoionized ADC, these and
all other elements will be more strongly ionized at relatively lower
temperatures. Correspondingly, the spectral lines and line ratios
predicted by our collisional equilibrium ADC models will certainly not
be representative of photoionized ADCs at similar temperatures.

Given our very simplified treatment of the ADC emission -- including
also the neglect of all radiative transfer effects (such as
[self-]absorption) -- we have not carried out any detailed
least-squares fits of disk+ADC models to the observed
spectra. Instead, we illustrate the type of match to the data that can
be achieved by adding an optically thin ADC component to the disk
models by means of just one example, shown in
Figure~\ref{fig-thinfits}. This shows that, in principle, the addition
of an optically thin ADC spectrum to an optically thick disk model
spectrum may well resolve both the UV color problem and the Balmer 
jump discrepancy for reasonable disk and ADC
parameters. More specifically, the model in Figure~\ref{fig-thinfits}
employs a disk radius that is consistent with the total eclipse width
(Section~\ref{general}), reasonable disk accretion rates which
correlate with the brightness change between August and November, ADC
parameters that are very similar to those derived by
Knigge~\&~Drew~(1997) and a distance to the system close to the most
recent estimate of Baptista~{et al.}~(1995). Since only a fraction of
the ADC will be occulted by the secondary star, it will also provide
an additional source of uneclipsed, optically thin radiation that may
help to account for the relatively high mid-eclipse flux levels and
for the absence of the Balmer jump in the optical mid-eclipse
spectra. How much of the ADC light should be expected to remain
uneclipsed will depend on the detailed geometry of and the run of
physical conditions within this region.

We stress that our disk+ADC models are only illustrative (since our
treatment of the ADC is so approximate) and that there are 
a number of questions that will have to be answered by modeling 
the combination of disk, wind and transition region in a more
self-consistent manner. For example, regardless of the wind 
driving mechanism, one might expect 
the wind mass loss rate (and hence the density in the ADC) to scale
approximately with the accretion rate through the disk
(e.g. \citeNP{livio4}). Correspondingly, we would have preferred to
scale the emission measures used in the disk+ADC models for the G160L
and PRISM data by the square of the ratio of the assumed accretion
rates in these fits. (In our constant density models, the appropriate
adjustment would have been to the ADC scale height, $H$.) However, we
found that matching the UV color in the G160L data always required an
emission measure similar to, or even greater than, that required to
smooth out the Balmer jump in disk+ADC model fits to the PRISM
data. In addition, the coronal equilibrium ADC models used in
Figure~\ref{fig-thinfits} do not produce any strong emission
lines in the observed wavelength range. Intuitively, and more in 
line with the results of Knigge \& Drew (1997), one might expect a
photoionized ADC to at least contribute to both the optical and UV 
lines that are observed.

\section{Conclusions}

We have presented and analyzed HST/FOS eclipse observations of the NL
CV UX~UMa. Two eclipses each were observed with the G160L grating in
August of 1994 and the PRISM in November of the same year. The main
results of our analysis are as follows:

\begin{trivlist}

\item[(1)] The system was approximately 50\% brighter in November than
in August. If interpreted as resulting from a change in accretion
rate, this indicates an increase in $\dot{M}_{acc}$ by $\gtappeq 50\%$
between the observations.

\item[(2)] From short to long wavelengths, UX~UMa's eclipses become
broader and shallower, and the (out-of-eclipse) flux level drops. 
These characteristics are consistent with the gradual occultation of
an accretion disk with a radially decreasing temperature
distribution.

\item[(3)] The light curves exhibit significant asymmetries about
orbital phase $\phi=0.0$ that are likely due to the BS at the
disk edge. The UV spectrum of the BS -- constructed as the pre- minus
post-eclipse difference spectrum -- exhibits absorption lines,
suggesting that part of the BS may be optically thin in the continuum
and veil the inner disk and/or the outflow from UX~UMa in these
lines. More direct evidence for the presence of optically thin
material in the BS region is provided by the fact that the Balmer jump
is strongly in emission in the optical BS spectrum derived from the
PRISM observations. 

\item[(4)] Disk models constructed as ensembles of SAs provide poor
descriptions of the observed accretion disk spectra if the latter are
identified with the spectra observed at orbital
phases away from eclipse. Compared to the data, disk model spectra are
too blue at UV wavelengths and exhibit excessively strong Balmer jumps
in the optical region. Surprisingly, suitably scaled, single temperature
model SAs with $T_{eff} \simeq 12,500-14,500$~K provide a
significantly better match to both the UV and optical data. Based on
this, we conclude that great care must be taken in attempts to derive
accretion rates from direct comparisons of disk models to
observations. This caution is relevant even to eclipse mapping
studies, since the transformation of surface brightness to disk
temperature {\em always} requires the specification of an emissivity
law. If the mismatch between disk models and CV spectra is due to 
``contamination'' of the uneclipsed light by additional, non-disk
radiation (see item [5] below), the problems just mentioned might be 
overcome by accounting explicitly for the additional spectral
component(s) when analyzing the data.

\item[(5)] One way to reconcile simple SA disk models with the
observed spectra is to postulate the presence of a significant amount
of optically thin gas in the system. The corresponding additional
spectral component would both flatten the UV spectral slope and fill
in the Balmer jump of the optically thick disk spectrum. The 
required optically thin material could be located in a
transition region between the disk photosphere and the fast wind from
the system. The existence of such an ``accretion disk chromosphere'' was
proposed by Knigge~\&~Drew~(1997) based on detailed modeling of HST
eclipse observations of the C~{\sc iv} wind line in UX~UMa. A simple
SA disk+ADC model that does not violate observational constraints and
uses ADC parameters similar to those suggested by
Knigge~\&~Drew~(1997)~does produce an improved, though still far from
perfect, fit to the observed UV and optical disk spectra. However, a
self-consistent calculation, in which photoionization and radiative
transfer effects are accounted for, will be needed to properly test
this idea. At least until then, other ways to resolve the discrepancy
between SA disk models and observed spectra (e.g. magnetic disk
truncation, better disk models) also remain viable.

\item[(6)] The presence of a multitude of uneclipsed emission lines in
the UV mid-eclipse spectra (some of which appear as absorption
features at out-of-eclipse phases) supports the suggestion made by
Knigge~{et al.}~(1997) that many, if not most, of the spectral lines
in the UV spectra of high-state non-magnetic CVs might arise in the
kind of ADC/wind region envisioned by Knigge~\&~Drew (1997).

\end{trivlist}

\acknow We gratefully acknowledge the support of NASA
through HST grant GO-5448 without which this work would not have been
possible. In addition, RAW acknowledges financial support from NASA
through grants NAGW-3171 and from STScI through grant GO-3683.03-91A,
both to the Pennsylvania State University. We would also like to thank
Chris Mauche for his contribution to this project and for a number of 
useful discussions related to this paper. We are also grateful to the 
referee for a very helpful report.

\bibliographystyle{apj}
\bibliography{bibliography}

\clearpage

\begin{deluxetable}{ll}
\tablewidth{250pt}
\footnotesize
\tablecaption{Adopted System Parameters (from Baptista~{et al.}
[1995])\label{tbl-systempars}}
\tablehead{
\colhead{Parameter}& 
\colhead{Value}}
\startdata
$q=M_2/M_1$ & $1.0 \pm 0.1$ \nl
$i$ & $71.0 \pm 0.6$ \nl
$M_1/M_{\sun}$ & $0.47 \pm 0.07$ \nl
$M_2/M_{\sun}$ & $0.47 \pm 0.1$ \nl
$R_1/R_{\sun}$ & $0.014 \pm 0.001$ \nl
$R_2/R_{\sun}$\tablenotemark{a}  & $0.53 \pm 0.04$ \nl
$a/R_{\sun}$ & $1.39 \pm 0.08$ \nl
\enddata
\tablenotetext{a}{The value listed for $R_2$, the volume-averaged radius of
the secondary star, is 4\% larger than that given by Baptista~{\em et
al.} (1995). The new value was derived using a more accurate formula for
$R_2/a$.}
\end{deluxetable}

\clearpage

\begin{deluxetable}{llcccccccc}
\tablewidth{500pt}
\footnotesize
\tablecaption{Journal of Observations\label{tbl-log}}
\tablehead{
\colhead{}& 
\colhead{}&
\colhead{UT}& 
\colhead{UT}&
\colhead{Phase Range}&
\colhead{No. of}& 
\colhead{}&
\colhead{}&
\colhead{}& 
\colhead{Wavelength} \nl
\colhead{}& 
\colhead{Date}&
\colhead{Start}&
\colhead{End}&
\colhead{(Cycles)}&
\colhead{Spectra}&
\colhead{Grating}&
\colhead{Detector}&
\colhead{Aperture}&
\colhead{Range (\AA)}}
\startdata
Run~1 & 8/3/94 & 02:53 & 03:55 & 0.91-1.13 & 691 & G160L & BLUE & {\bf 1.0} & 1140-2508 \nl
Run~2 & 8/3/94 & 07:43 & 08:45 & 0.94-1.16 & 691 & G160L & BLUE & {\bf 1.0} & 1140-2508 \nl
Run~3 & 11/11/94 & 04:34 & 05:52 & 0.84-1.11 & 870 & PRISM &RED & {\bf 4.3} & 1850-8950 \nl
Run~4 & 11/11/94 & 18:55 & 20:13 & 0.88-1.16 & 870 & PRISM &RED & {\bf 4.3} & 1850-8950 \nl
\enddata
\end{deluxetable}

\clearpage

\begin{deluxetable}{llllll}
\tablewidth{450pt}
\footnotesize
\tablecaption{New eclipse timings of UX~UMa with respect to the
ephemeris of Baptista~{et al.} (1995)\label{tbl-timings}}
\tablehead{
\colhead{}& 
\colhead{Cycle Number}& 
\colhead{HJD-2,440,000\tablenotemark{a}}&
\colhead{O-C (Cycles)\tablenotemark{a}}&
\colhead{HJD-2,440,000\tablenotemark{b}}&
\colhead{O-C (Cycles)\tablenotemark{b}}}
\startdata
Run 1 & 28793 & 9567.6353 & +0.0025 & 9567.6353 & +0.0026 \nl
Run 2 & 28794 & 9567.8323 & +0.0040 & 9567.8319 & +0.0021 \nl
Run 3 & 29368 & 9680.7215 & +0.0034 & 9680.7213 & +0.0023 \nl
Run 4 & 29371 & 9681.3115 & +0.0033 & 9681.3113 & +0.0025 \nl
\enddata
\tablenotetext{a}{Measured from parabolic fit to eclipse center.}
\tablenotetext{a}{Measured from derivative light curve.}
\end{deluxetable}

\clearpage

\figcaption[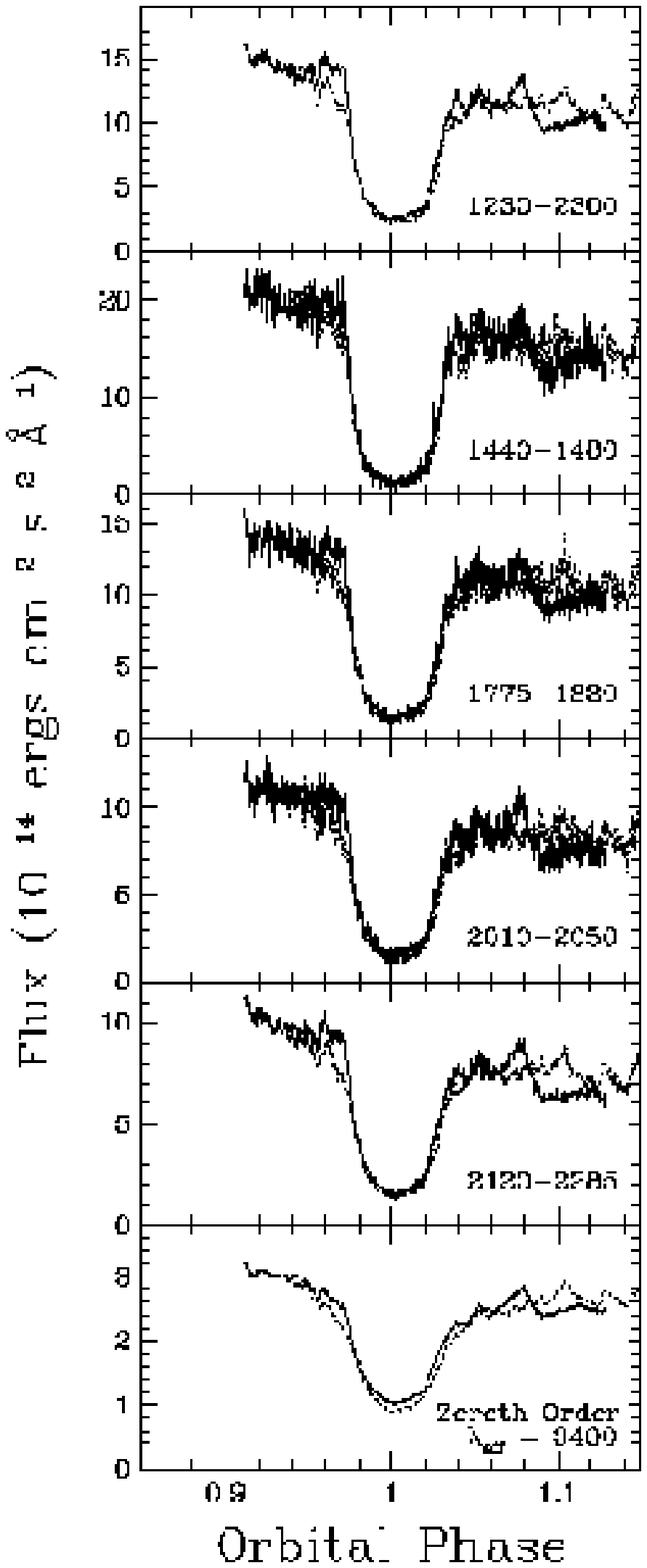]{Continuum light curves constructed
from the August (G160L) observing sequences (Run 1 -- solid lines;  Run 2
-- dotted lines) phased according to the ephemeris of Baptista~{\em
et al.} (1995). The wavelength ranges over which the spectra were
averaged to obtain the light curves are given (in~\AA ) at the
bottom right of each panel; for the zeroth order light curves the
effective wavelength of the bandpass is given.\label{fig-g160lcontlights}}

\figcaption[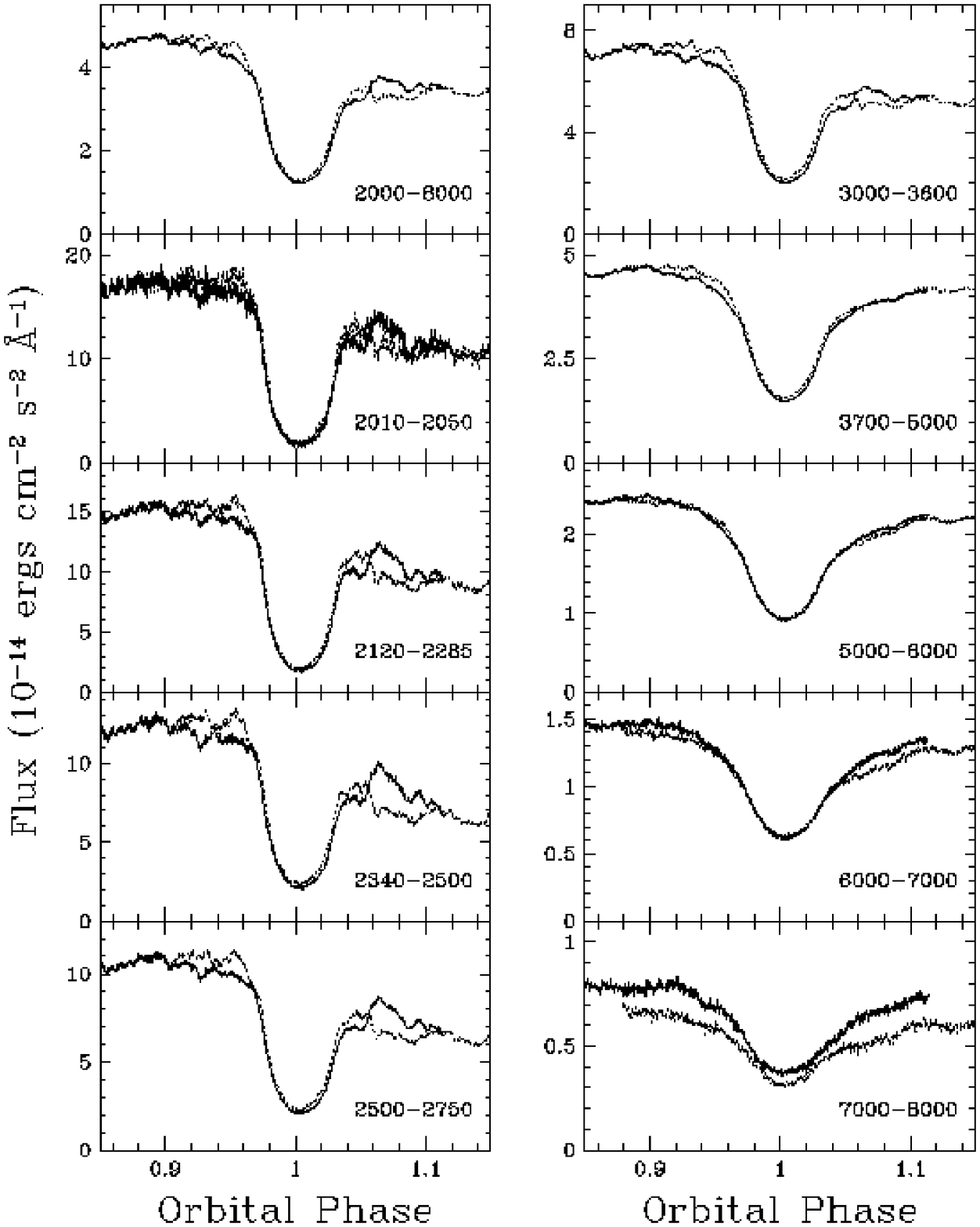]{Continuum light curves constructed
from the November (PRISM) observing sequences (Run 3 -- solid lines;
Run 4 -- dotted lines) phased according to the ephemeris of
Baptista~{et al.} (1995). The wavelength ranges over which the
spectra were averaged to obtain the light curves are given (in \AA ) 
at the bottom right of each panel.\label{fig-prismcontlights}}

\figcaption[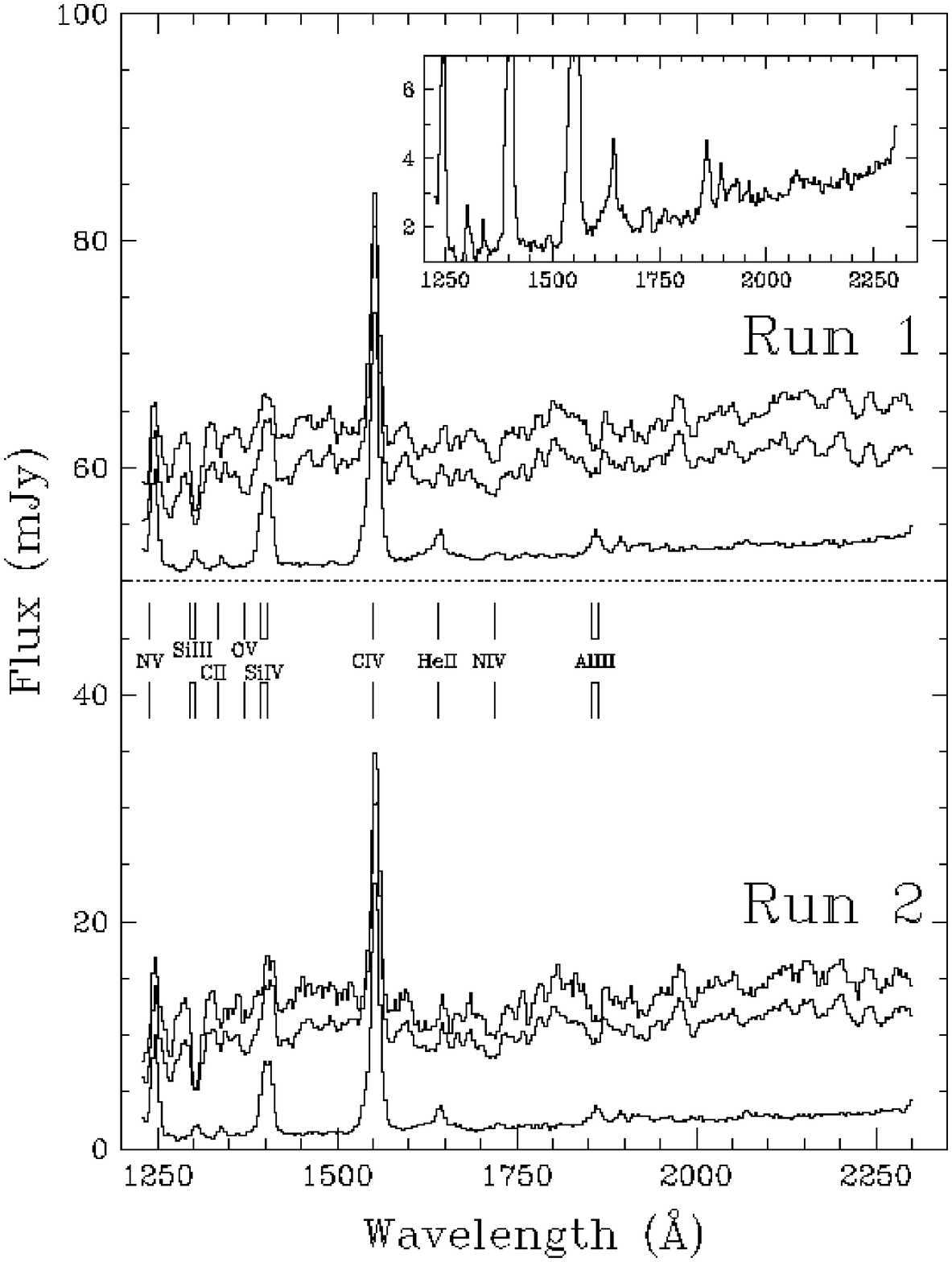]{The average pre-, mid- 
and post-eclipse spectra for Runs~1 and 2. The dotted horizontal line
marks the zero level for the Run~1 spectra. The inset shows the Run~1
mid-eclipse spectrum on a scale more appropriate for inspection
of the weaker emission lines in this spectrum. For both runs, the 
highest (middle, lowest) spectrum corresponds to the average over 
pre- (post-, mid-) eclipse orbital phases.
\label{fig-g160lavespecs}}

\figcaption[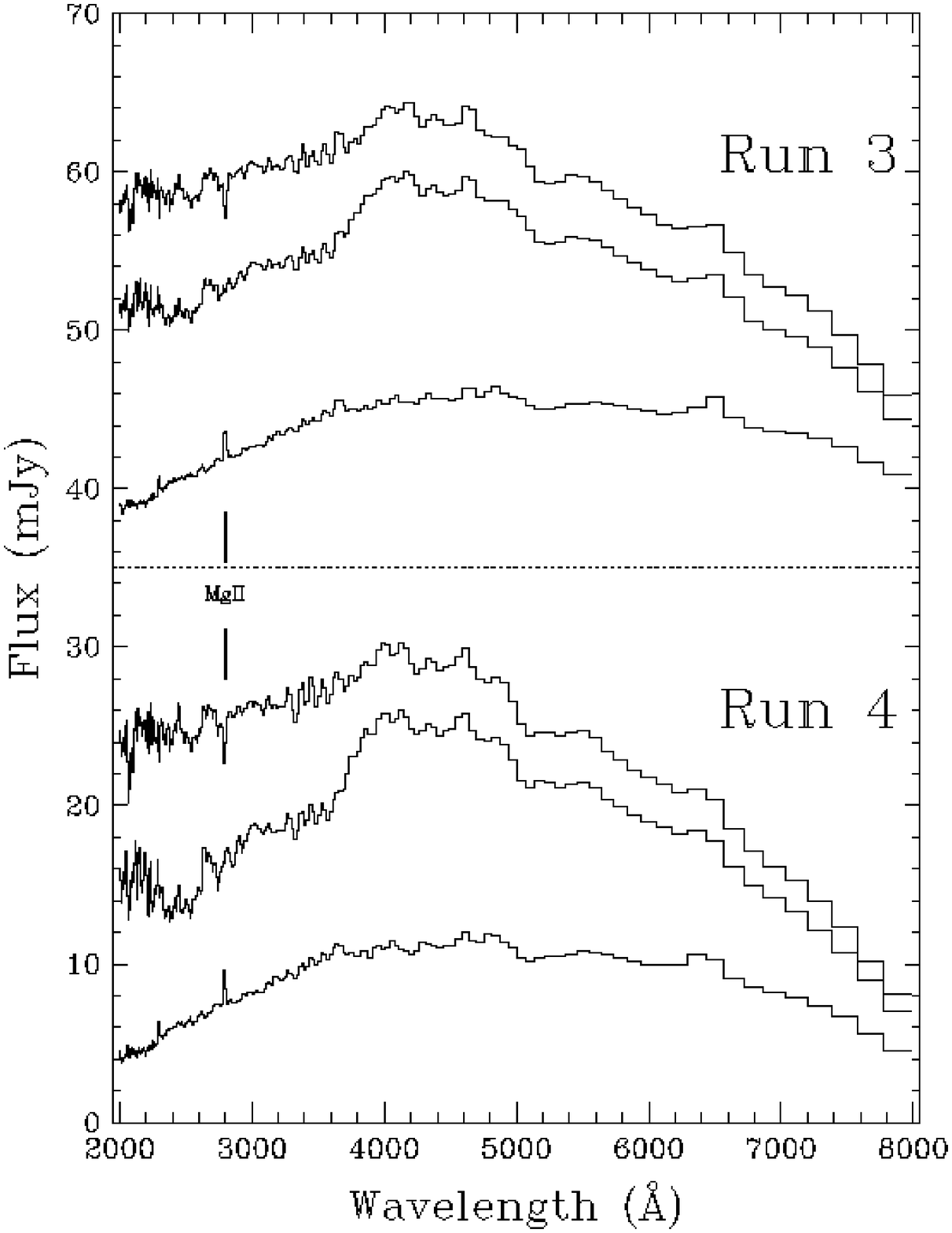]{The average pre-, mid- and 
post-eclipse spectra for Runs~3 and 4. The dotted horizontal line
marks the zero level for the Run~3 spectra. For both runs, the 
highest (middle, lowest) spectrum corresponds to the average over 
pre- (post-, mid-) eclipse orbital phases. \label{fig-prismavespecs}}

\figcaption[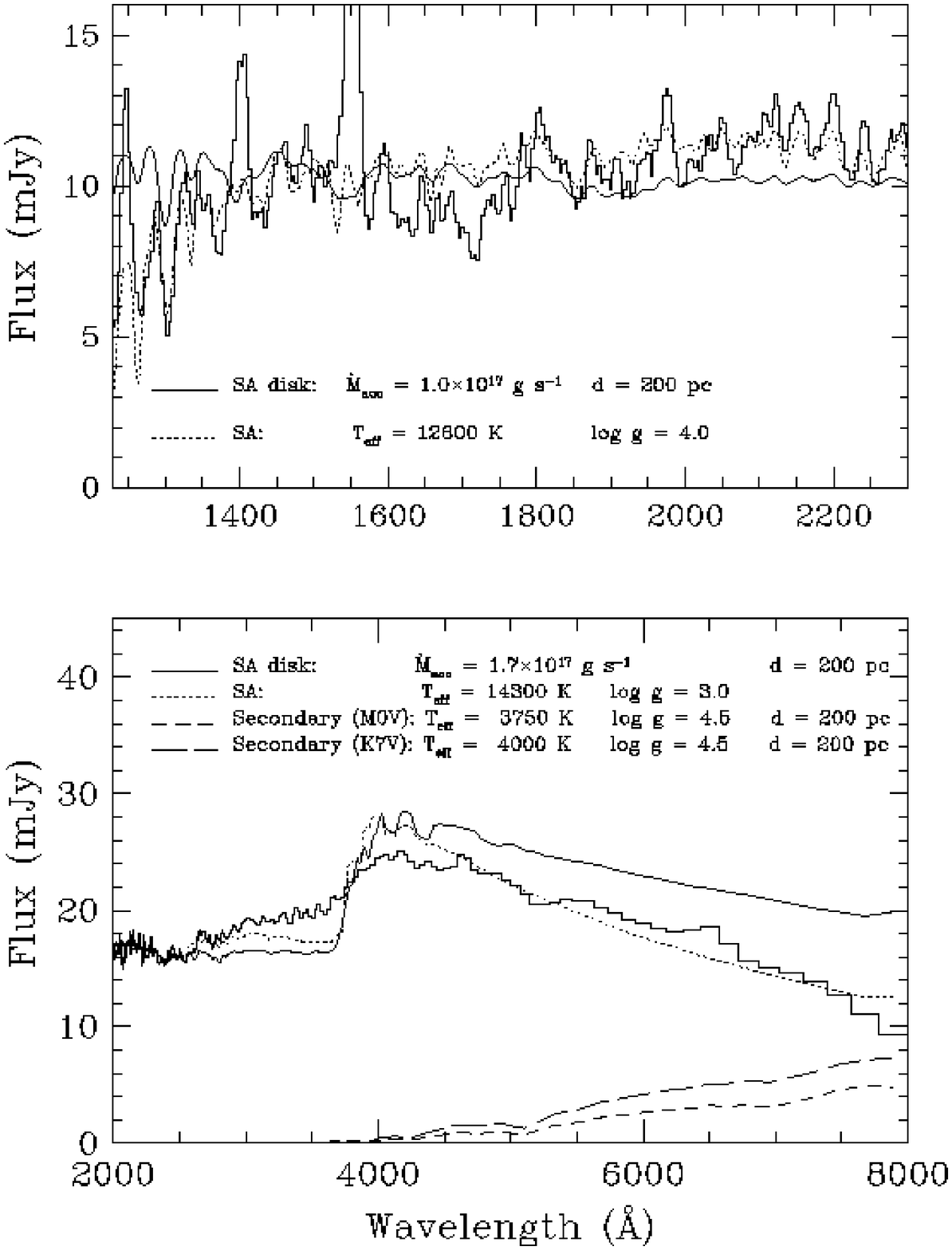]{Model fits to observed Run~1 G160L (top
panel) and Run~3 PRISM (bottom panel) post-eclipse spectra (shown as
histograms). In both panels, the solid and dotted lines are the
best-fitting disk model and single temperature model stellar 
atmosphere, respectively. For a distance towards UX~UMa of 345~pc, 
the normalization of the stellar atmosphere models would require projected
emitting areas with a characteristic minimum linear size
of 15 white dwarf radii for the fit to the G160L 
data and 14 white dwarf radii for that to the PRISM
data. In the bottom panel, two model stellar atmospheres with spectral
types likely to bracket that of the secondary star are also plotted 
(short- and long-dashed lines). These have been scaled according to
the apparent secondary star radius (see text) and the 200~pc distance
implied by the best-fitting disk model.\label{fig-postfits}}

\figcaption[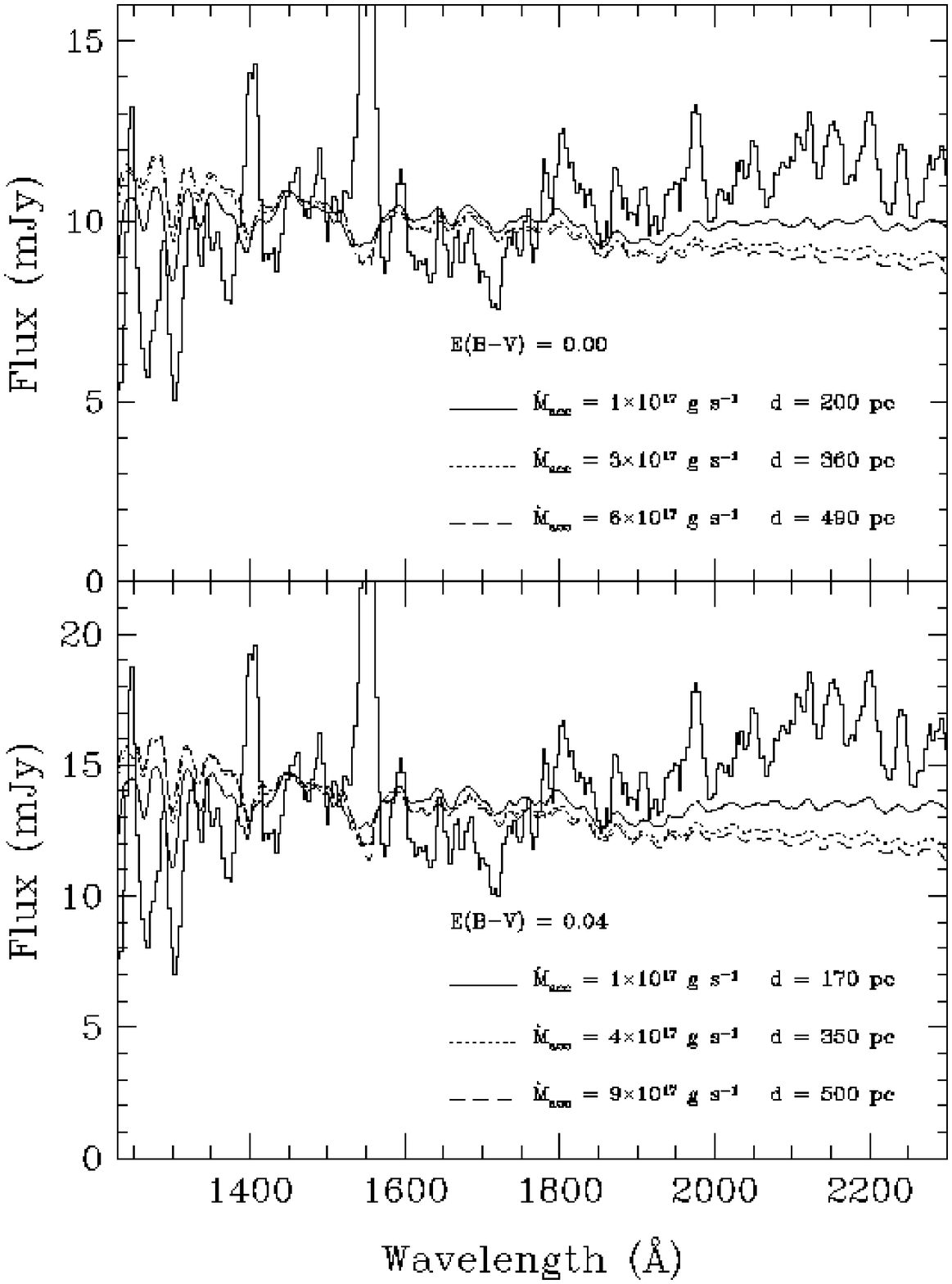]{Comparison of the original (top panel) 
and maximally dereddened (bottom panel) Run~1 G160L post-eclipse spectra 
against disk models with varying accretion rates. All models 
have been normalized to the observed flux at 1450~\AA~. The accretion rates of
the three models shown in each panel have been chosen so as to make
the implied distances for this choice of normalization cover the
allowed distance range of 200~-~500~pc.\label{fig-dred1fits}} 

\figcaption[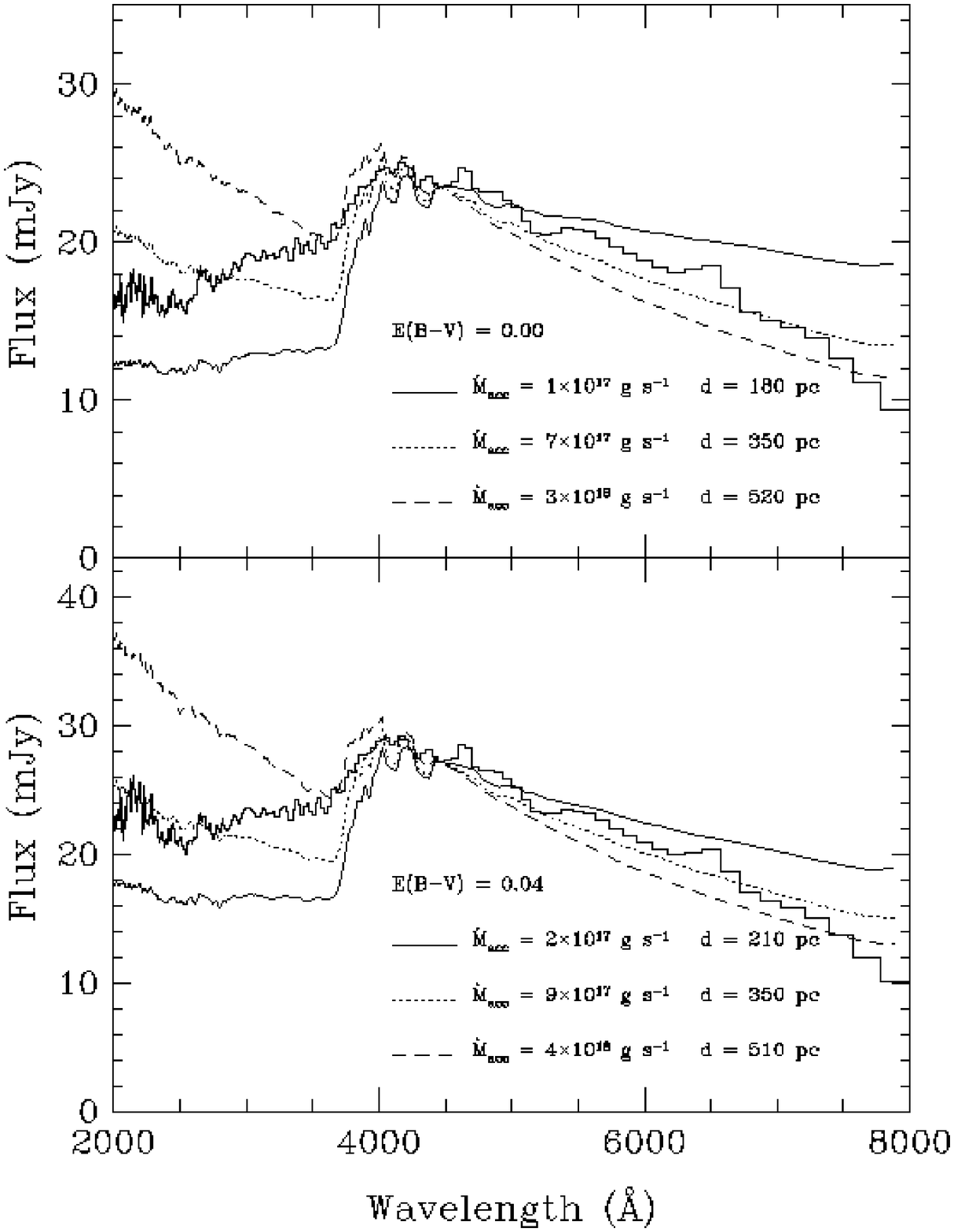]{Comparison of the original (top panel)
and maximally dereddened (bottom panel) Run~3 PRISM post-eclipse spectra
against disk models with varying accretion rates. All models have been
normalized to the observed flux at 4500~\AA~. The accretion rates of the three
models shown in each panel have been chosen so as to make the implied 
distances for this choice of normalization completely cover the
allowed distance range of 200~-~500~pc.\label{fig-dred3fits}}

\figcaption[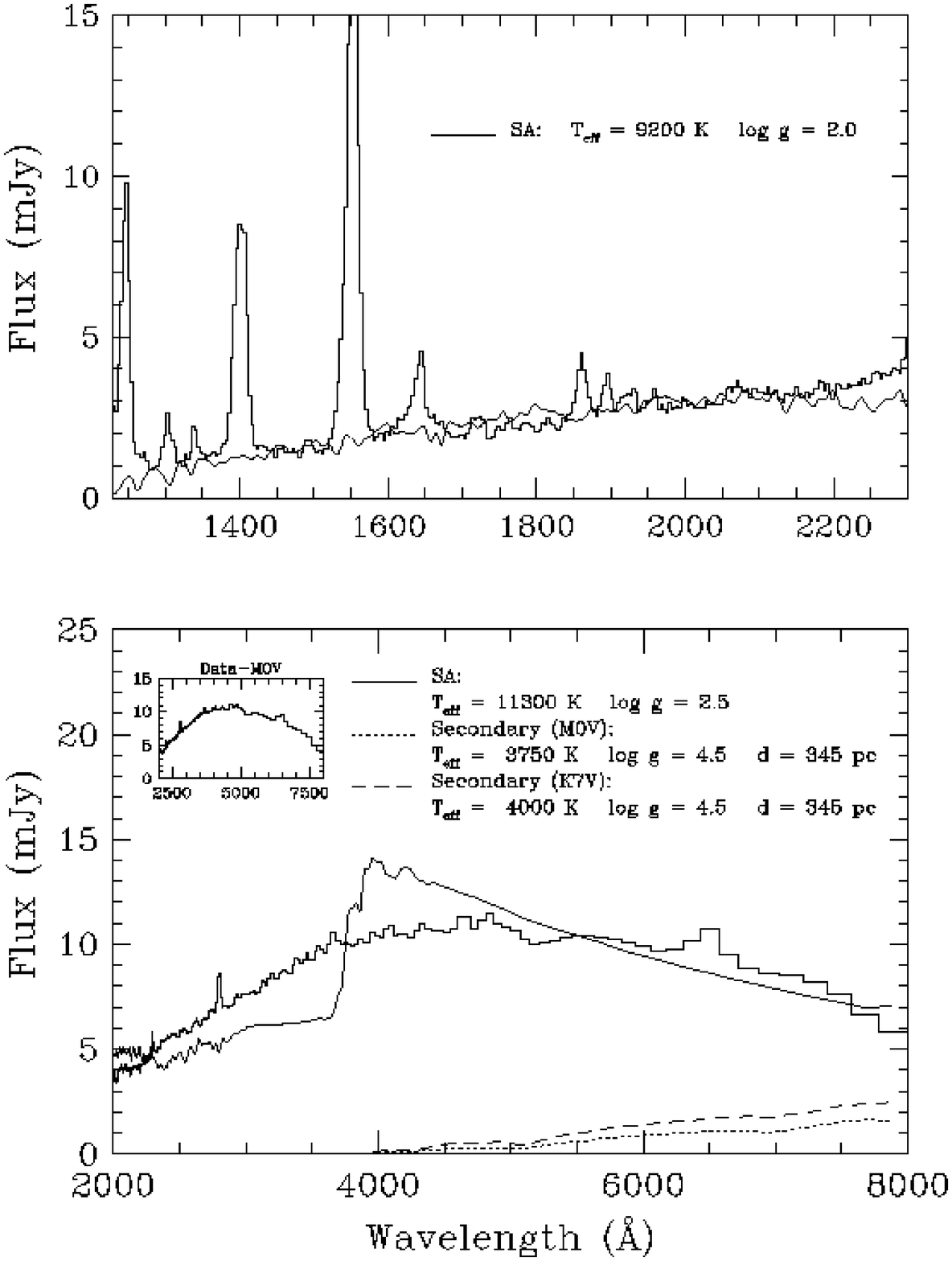]{Model fits to Run~1 G160L (top
panel) and Run~3 PRISM (bottom panel) mid-eclipse spectra (shown as
histograms). In each panel, the solid line corresponds to the
best-fitting single temperature model stellar atmosphere. 
For a distance towards UX~UMa of 345~pc, the normalization of the 
models requires projected emitting areas with minimum linear sizes of
$l_{min}=20~R_{WD}$ (G160L) and $l_{min}=12~R_{WD}$ (PRISM).
In the bottom panel, two model stellar atmospheres with spectral
types likely to bracket that of the secondary star are also plotted 
(dotted and dashed lines). These have been scaled according to 
the apparent secondary star radius (see text) and a 
distance of 345~pc. The inset in the bottom panel shows the result of
subtracting the M0V main-sequence star spectrum from the 
data.\label{fig-midfits}}

\figcaption[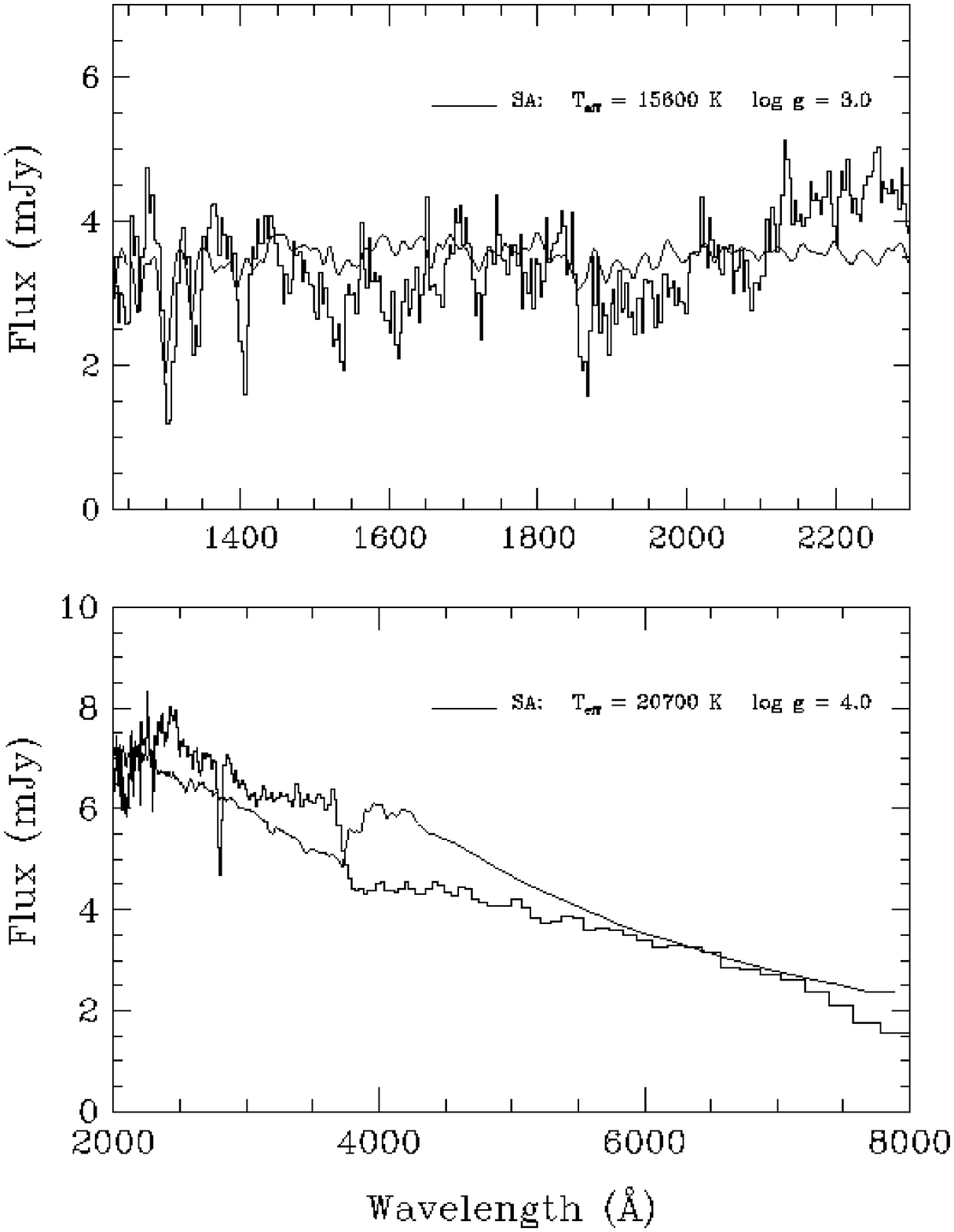]{Model fits to the bright spot
spectra constructed from the Run~1 G160L data (top panel) and the
Run~3 PRISM data (bottom panel). Bright 
spot spectra have been calculated for each observing sequence by
subtracting the average post-eclipse spectrum from the average
pre-eclipse spectrum. In each panel, the data are plotted as a
histogram and the solid line corresponds to the
best-fitting single temperature model stellar atmosphere. 
For a distance towards UX~UMa of 345~pc, the normalization of the 
models would require projected emitting areas with minimum linear
sizes of $l_{min}=5~R_{WD}$ in both cases.\label{fig-spotfits}}

\figcaption[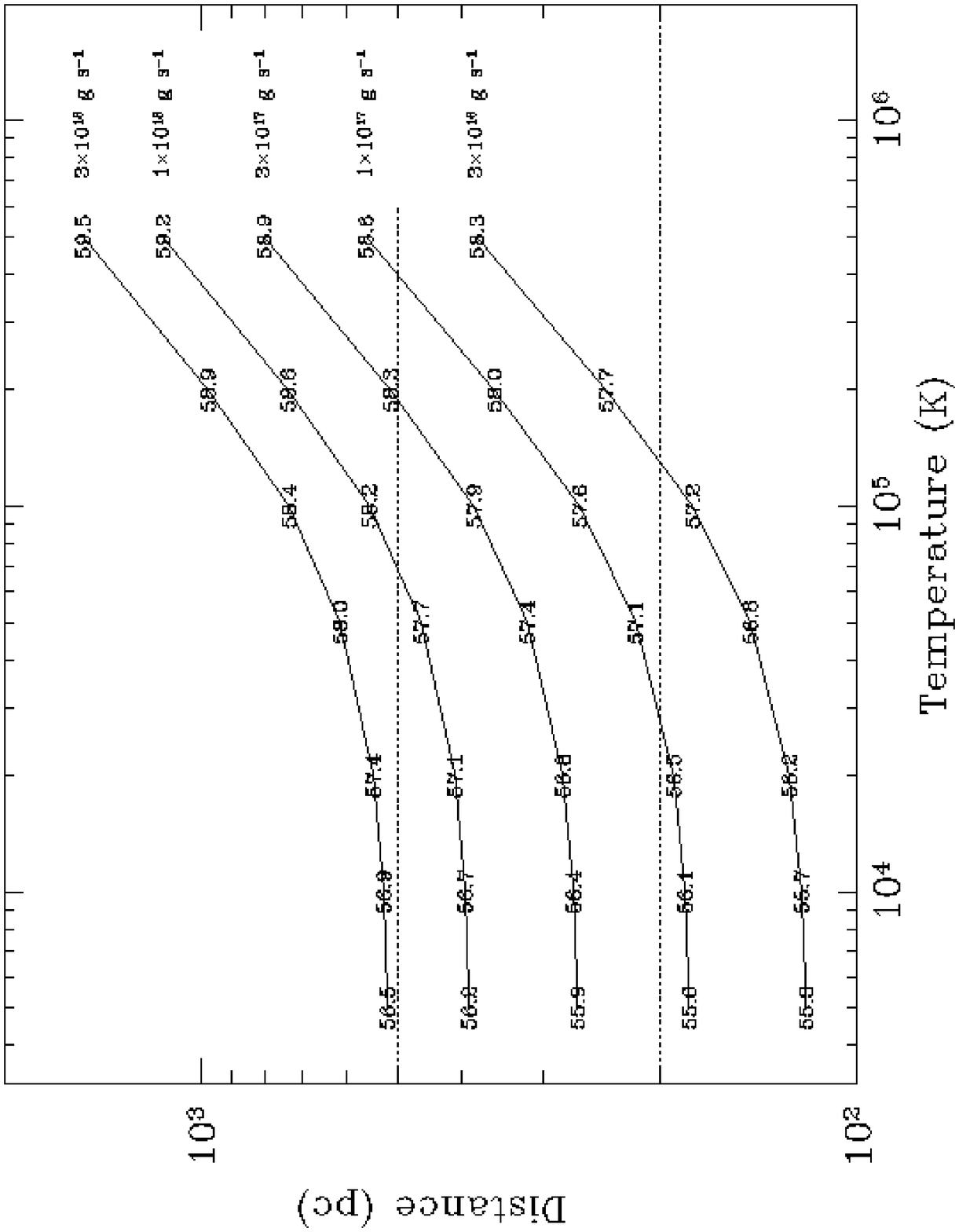]{Emission measures required to fill in the
Balmer jump in disk model spectra. Each connected sequence of numbers
in this figure gives the emission measures needed to completely fill in
the Balmer jump in the spectrum of a disk model with given accretion
rate by adding an optically thin H~{\sc i} recombination continuum
component. The emission measures are placed on the plot according to
the electron temperature of the optically thin component (x-axis) and
the implied distance (y-axis) if the combined disk+recombination model
is to match the Run~3 post-eclipse spectrum at 4500~\AA. The
horizontal dotted lines mark the upper and lower limits of 200~pc and
500~pc, respectively, on the distance towards
UX~UMa.\label{fig-thinstuff1}}

\figcaption[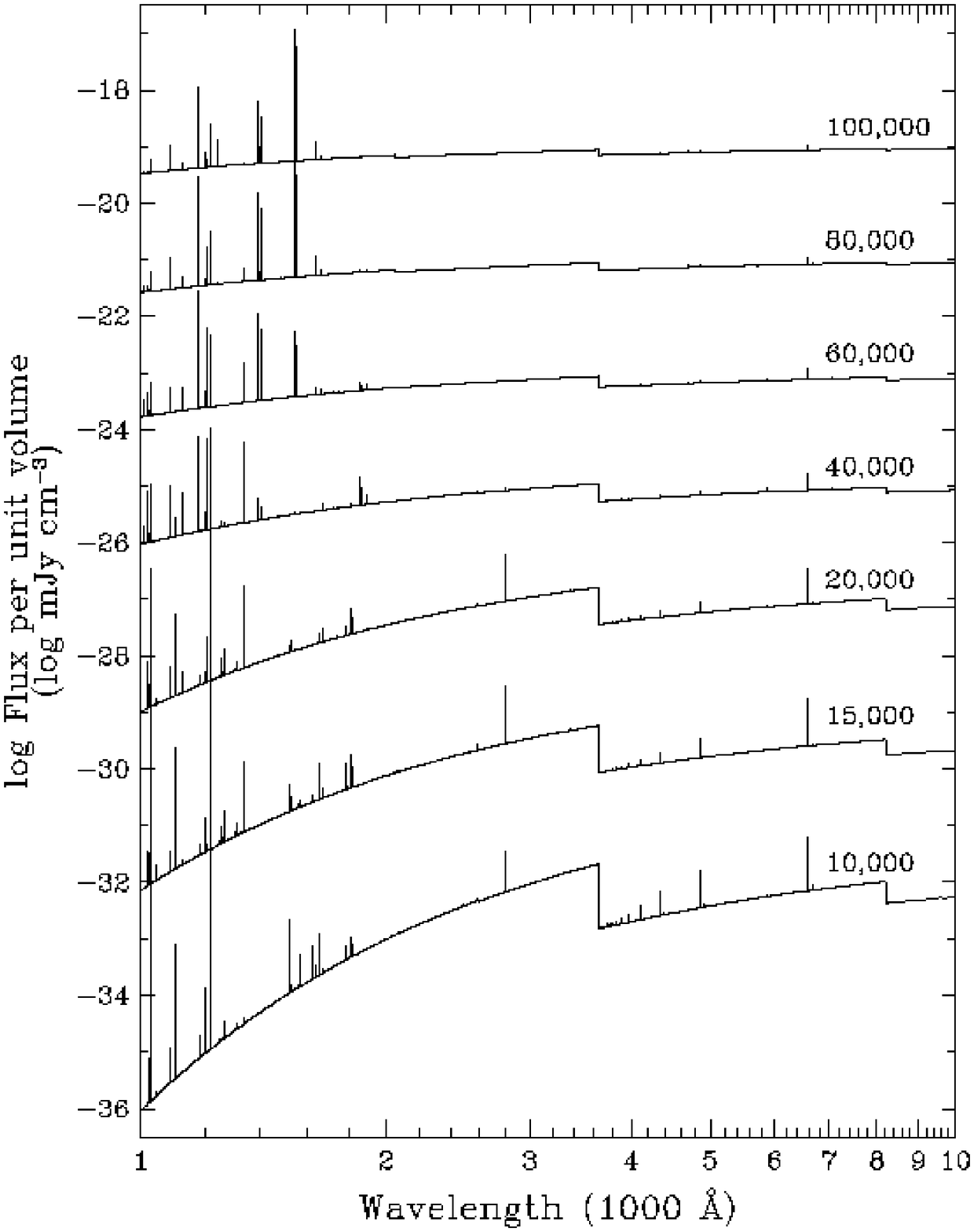]{The spectra emitted by 1~cm$^{-3}$ of solar
abundance plasmas in coronal (collisional) equilibrium are shown for a
variety of electron temperatures. The lowest temperature spectrum
is shown on the true scale; all others have been displaced vertically
by successive factors of 100.\label{fig-coronal}}

\figcaption[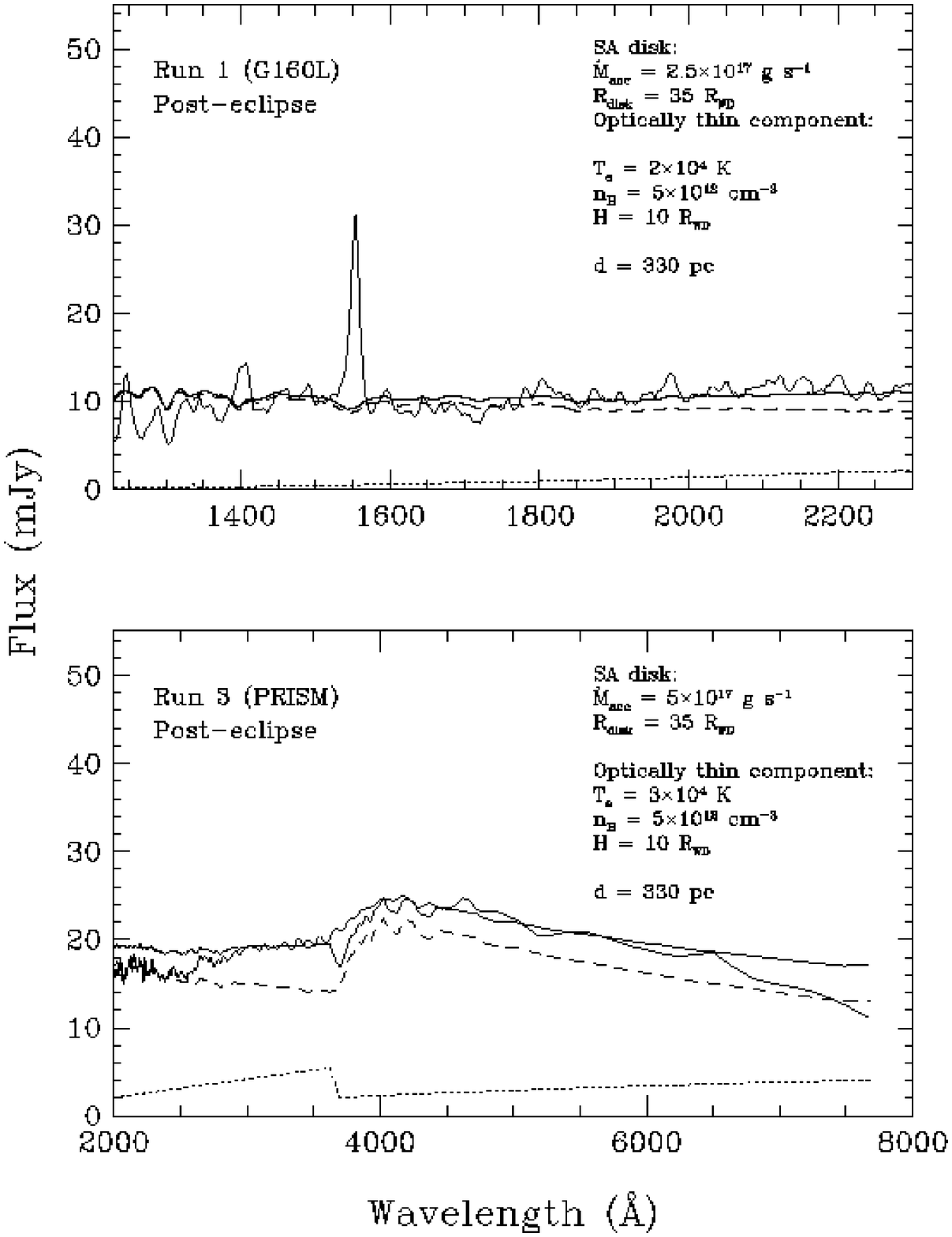]{Illustrative fits to the Run~1 (top panel)
and 3 (bottom panel) post-eclipse spectra, shown as thin solid
lines. The models (thick solid lines) are combinations of optically 
thin recombination spectra (dotted lines) and SA disk models (dashed
lines). The parameters adopted for the fits are noted in the figures;
the volume of the optically thin gas is $V=\pi R_{disk}^{2}H$, so if 
the optically thin material is located in a roughly cylindrical
region above the surface of the accretion disk, $H$ is the
corresponding vertical scale height.\label{fig-thinfits}}

\clearpage

\begin{figure}
\figurenum{Figure 1}
\plotone{f1.ps}
\end{figure}

\clearpage 

\begin{figure}
\figurenum{Figure 2}
\plotone{f2.ps}
\end{figure}

\clearpage 

\begin{figure}
\figurenum{Figure 3}
\plotone{f3.ps}
\end{figure}

\clearpage

\begin{figure}
\figurenum{Figure 4}
\plotone{f4.ps}
\end{figure}

\clearpage

\begin{figure}
\figurenum{Figure 5}
\plotone{f5.ps}
\end{figure}

\clearpage

\begin{figure}
\figurenum{Figure 6}
\plotone{f6.ps}
\end{figure}

\clearpage

\begin{figure}
\figurenum{Figure 7}
\plotone{f7.ps}
\end{figure}

\clearpage

\begin{figure}
\figurenum{Figure 8}
\plotone{f8.ps}
\end{figure}

\clearpage

\begin{figure}
\figurenum{Figure 9}
\plotone{f9.ps}
\end{figure}

\clearpage

\begin{figure}
\figurenum{Figure 10}
\plotone{f10.ps}
\end{figure}

\clearpage

\begin{figure}
\figurenum{Figure 11}
\plotone{f11.ps}
\end{figure}

\clearpage

\begin{figure}
\figurenum{Figure 12}
\plotone{f12.ps}
\end{figure}

\end{document}